\documentclass[11pt]{article} 
\usepackage[utf8]{inputenc}
\usepackage{multirow}
\usepackage{amsfonts}
\usepackage{amsmath}
\usepackage{amssymb}
\usepackage{amsthm}
\usepackage{caption}
\usepackage[dvipsnames]{xcolor}
    \definecolor{darkgreen}{rgb}{0,0.5,0}
    \definecolor{darkblue}{rgb}{0,0,0.6}
    \definecolor{purple}{rgb}{0.4,.2,0.7}
\usepackage[margin = 2.5cm]{geometry}
\usepackage{graphicx}
\usepackage[hyperfootnotes = false, colorlinks = true, linkcolor = darkblue, citecolor = purple]{hyperref}
\usepackage{subcaption}
\usepackage{ytableau}

\newcommand{\be}{\begin{equation}}
\newcommand{\ee}{\end{equation}}
\newcommand{\bea}{\begin{eqnarray}}
\newcommand{\eea}{\end{eqnarray}}
\def\la{\label}
\def\nref#1{(\ref{#1})}
\def\half{{1 \over 2 }}

	\newcommand{\bes}{\begin{equation} \begin{split} }	
	\newcommand{\ees}{\end{split} \end{equation} }

	\newcommand{\lp}{\left (}
	\newcommand{\rp}{\right )}

	\newcommand{\ra}{\rightarrow}

	\usepackage{physics}

\def\etanew{{\bar \theta}}

\begin{document}

\thispagestyle{empty}
\begin{center}
    ~\vspace{5mm}

  {\LARGE \bf {Scaling similarities and quasinormal modes of D0 black hole solutions \\}}

 %   {\bf   Juan Maldacena$^1$}

     %   $^1$Institute for Advanced Study,  Princeton, NJ 08540, USA 

   \vspace{0.5in}
     
   {\bf    Anna Biggs$^1$ and  Juan Maldacena$^2$ 
   }

    \vspace{0.5in}

  $^1$
  Jadwin Hall, Princeton University,  Princeton, NJ 08540, USA 
   \\
   ~
   \\
  $^2$
  Institute for Advanced Study,  Princeton, NJ 08540, USA

    \vspace{0.5in}

    \vspace{0.5in}
    
%     {\tt   malda@ias.edu}

\end{center}

\vspace{0.5in}

\begin{abstract}
 
 We study the gravity solution dual to the D0 brane  quantum mechanics, or BFSS matrix model, in the 't Hooft limit.
 
 The classical physics described by this gravity solution is invariant under a scaling transformation, which changes the action  with a specific critical exponent, sometimes   called the hyperscaling violating exponent. We present an argument for this critical exponent from the matrix model side, which leads to an explanation for the peculiar temperature dependence of the entropy in this theory, $S \propto T^{9/5}$. We also present a similar argument for all other $Dp$-brane geometries. 

We then compute the black hole quasinormal modes. This   involves perturbing the finite temperature geometry. These perturbations can be easily obtained by a mathematical trick where we view the solution as the  dimensional reduction of an $AdS_{ 2 + 9/5 } \times S^8$ geometry.
  
 \end{abstract}
 
\vspace{1in}

\pagebreak

\setcounter{tocdepth}{3}
{\hypersetup{linkcolor=black}\tableofcontents}

 \section{Introduction and Motivation}

 The duality between the BFSS \cite{Banks:1996vh} matrix model and the D0 brane near horizon geometry is one of the simplest examples of holographic dualities. 
 %Namely, it is a duality between a quantum mechanical system and a gravitational solutions with certain asymptotic boundary conditions. 
 This is a rich system with different energy regimes \cite{Itzhaki:1998dd}. In this paper, we will concentrate on the first interesting regime as we go down in energies. Namely, we consider the large $N$ theory in the range of temperatures 
 \be \la{TemRa}
  1 \ll N   ~,\quad \quad N^{-10/21} \ll   {T \over \lambda^{1/3} } \ll 1~,\quad \quad \lambda = g^2 N 
  \ee 
  where $g^2$ is the overall coupling playing the role of $\hbar$ in the Matrix quantum mechanics and has dimensions of (energy)$^3$.  Here $\lambda^{1/3}$ is an energy scale that sets the effective coupling of the matrix model (also called the 't Hooft coupling). In this regime, the matrix model is strongly coupled and the system is described by the near horizon geometry of a ten dimensional charged black hole, whose precise geometry we will discuss later. 
 
      This is a simple example of a quantum mechanics system (as opposed to a quantum field theory) which has a holographic dual described by Einstein gravity (as opposed to a more exotic gravity theory). In the temperature range  \nref{TemRa},  this quantum many-body system is displaying a quantum critical behavior which we will call a ``scaling similarity" \cite{Boonstra:1998mp}. The geometry is such that under a time rescaling $t \to \gamma t$ and a suitable rescaling of the radial dimension, the action changes by $S \to \gamma^{ -\etanew } S$, with $\etanew$ a certain scaling exponent. Since the action changes, this is not a symmetry of the quantum theory. However, at large $N$ we can just do classical physics in the bulk, and this transformation becomes an actual symmetry of the equations of motion. We describe this symmetry in detail and compute the scaling exponent $\etanew$ from the matrix model side, see also \cite{Smilga:2008bt,Wiseman:2013cda,Morita:2013wfa}. What we call a ``scaling similarity" has also been called hyperscaling violation \cite{Fisher:1986zz,Huijse:2011ef}, and the exponent $\etanew = - \theta$, where $\theta $ is the hyperscaling violating exponent defined in \cite{Fisher:1986zz,Huijse:2011ef}\footnote{ We find it convenient to define it with the opposite sign because $\etanew >0$ for the main cases we will consider. }. Other $Dp$ brane metrics display a similar critical behavior \cite{Boonstra:1998mp,Dong:2012se}.   We also explain how this scaling similarity can be used to classify the perturbations propagating on the geometry, which are dual to operators in the quantum theory. 
  
  An interesting property of this strongly coupled many-body system is its response to perturbations. In the bulk, these are described by black hole excitations called  quasinormal modes \cite{Horowitz:1999jd}, see \cite{Berti:2009kk} for a review.  Each quasinormal mode decays with a particular complex frequency proportional to the black hole temperature. Their imaginary part arises because the perturbations fall into the black hole. This is viewed as thermalization in the quantum system.   Notably, even though the system has $N^2$ degrees of freedom, the quasinormal mode frequencies are independent of $N$. In addition, the  number of modes below any given frequency is also independent of $N$. This is an important feature of any quantum system that has a black hole description within Einstein gravity.

    The quasinormal modes are the fingerprint of the black hole.  Their frequencies are fixed by the geometry around the black hole and are independent of the initial perturbation. So they are among the definite predictions that the duality makes. Namely, a simulation of this model in either a classical or quantum computer should reproduce these modes. Finding them would constitute evidence that the quantum system gives rise to black holes. In fact, we could say it is evidence similar in spirit to the evidence we have from LIGO observations \cite{Ghosh:2021mrv}---we are observing some vibrations whose waveforms (or frequencies) are computed by solving Einstein's equations.

    The computation of quasinormal modes involves expanding all fields to quadratic order around the background\footnote{After submission we learnt that the quasinormal mode problem for these black holes had been already considered in \cite{Craps:2016cgo}, where the two lowest SO(9) invariant QNM solutions were found.}. This seems a somewhat daunting task. However, we are helped by a few tricks. First we note that the solution itself can be formally viewed as the dimensional reduction from an $AdS_{2+\etanew} \times S^8$ solution in $10+\etanew$ dimensions where $\etanew$ is a fractional number equal to the action scaling exponent mentioned above \cite{Kanitscheider:2009as}. Then, we can use the spectrum of dimensions of operators derived in \cite{Sekino:1999av}  to determine the masses for the fields. In fact, we will also mention a quick trick to get the spectrum by going to eleven dimensions.  In addition,   the finite temperature solution is the usual $AdS_{2+\etanew}$ black brane. Then we can simply write the wave equation for a massive scalar field on this background. Solving this equation numerically we find the quasinormal modes. 
   
   As a side remark, in appendix \ref{RFour}, we provide a quick rederivation of the one loop eight derivative correction to the thermodynamics of the black hole which was originally obtained in \cite{Hyakutake:2013vwa,Hyakutake:2014maa}. 
   
 \section{The D0 brane gravity background } 
  
 The gravity solution dual to the matrix model, in the regime \nref{TemRa} is 
 \bea \la{SolIIA}   { ds^2_{str} \over \alpha'} &=& \rho^{-3/2} \left( - \rho^5 h  d\tau^2 + { d \rho^2 \over h \rho^2 } + d\Omega_8^2 \right)
 \cr 
 h & =& 1 - { \rho_0^7 \over \rho^7} 
 \cr 
 e^{ \phi} &=& { (2 \pi)^2 \over d_0 N }   \rho^{ -21/4}   ~,~~~~~~~~~~~~~ 
   A =\sqrt{\alpha'} {d_0 N \over (2 \pi)^2 } \rho^7 d\tau ~,~~~~~~~~~~~~~d_0 = 240 \pi^5 
   \cr 
   \tau &=& ( d_0 \lambda)^{1/3} t ~,~~~~~~\lambda \equiv  g^2 N 
 ~,~~~~~~T_\tau = { 7 \over 4 \pi } \rho_0^{5/2} ~,~~~~~~T=T_t = (d_0 \lambda)^{1/3} T_\tau  
  \eea
 where we have written the metric  in string frame. The time $t$ should be identified with the time in the matrix model. We see that the dimensionful coupling $\lambda $ is only setting the units of time. In other words, in terms  of the dimensionless time $\tau$ the metric is completely independent of $\lambda$.  We have given the temperature both with respect to the dimensionless time $\tau$, $T_\tau$,  and the matrix model time $t$, $T$. The solution is valid in the following range of temperatures
 \be     N ^{ -10/21 } \ll   {T \over  \lambda^{1/3} }  \ll 1 %~,~~~~{\rm and} ~~~~~N \gg 1
\ee 
The upper limit comes from demanding that the curvature of the sphere is not too large at the horizon. This radius of curvature  becomes of order one in string units  when $\rho_{0} \sim 1$. The lower limit comes when the dilaton becomes of order one, or $ e^\phi|_{\rho = \rho_0} \sim 1$.    Of course, for this range to be wide enough, we need $N\gg 1$. In our conventions, $16 \pi G_N = (2 \pi)^7 \alpha'^4$, and $\alpha'$ disappears from the action. 
The metric can also be written as 
\be \la{AdSCo}
ds^2 \propto z^{3/5} \left[  \left({ 2 \over 5} \right)^2 \left( { - h d\tau^2 + d z^2/h \over z^2 }\right) + d\Omega_8^2 \right]~,~~~~~~~~h = 1 - \left( { z  \over z_0  } \right)^{14/5} ~,~~~~~~~~ z = { 2 \over 5 } \rho^{ -5/2} 
 \ee 
 which shows that, for $h=1$,  the geometry is conformally equivalent to $AdS_2 \times S_8$ \cite{Boonstra:1998mp,Skenderis:1998dq,Kanitscheider:2008kd}.  

 Strictly speaking,  the thermodynamic state described by \nref{SolIIA} is not quite stable, since it can decay by emission of D0 branes that go to infinity. In the large $N$ limit this process is exponentially suppressed (see the rates computed  in e.g. \cite{Lin:2013jra}) and we will ignore it in this paper.% \JM{Added comment on emission of D0 branes}
 
 \section{Similarity transformations }

 An important property of this solution is that under the transformation 
 \be \la{ResCa}
  t \to \gamma t ~,\quad \quad z \to \gamma z  ~,~~( {\rm or} ~~ \rho \to \gamma^{-2/5} \rho )  ~,\quad \quad N \to N ~,\quad \quad \lambda \to \lambda 
  \ee 
  the metric gets rescaled, and the dilaton gets shifted in such a way that the whole action is changed by
  \be \la{ActExp}
  S \to \gamma^{-\etanew    } S ~,~~~~~~~\etanew   = { 9 \over 5 } 
  \ee 
 This is in contrast to usual purely $AdS$ solutions where the rescaling \nref{ResCa} is an actual symmetry of the metric that leaves the action invariant. This means that the transformation \nref{ResCa} is a symmetry of the equations of motion and of classical observables. But it is not a symmetry of the quantum theory. However, if we are interested in   leading order in $N$  results, the classical theory is enough, and this is as good as a symmetry. 
 
 Transformations that rescale the action are sometimes called ``similarities'' \cite{LandauL} rather than symmetries, and that is the name we will use here.  This type of similarities have also been called ``hyperscaling violation'', with the hyperscaling violating exponent $\theta = - \etanew  $ \cite{Fisher:1986zz,Huijse:2011ef}.

Similarities are common. For example, Einstein gravity with zero cosmological constant   has a well known similarity under $g_{\mu \nu} \to \kappa^2 g_{\mu\nu}$ under which the action changes as $S \to \kappa^{D-2} S$ .  This can be used to argue that the Euclidean action  of a Schwarzschild black hole in $D$ spacetime dimensions should scale like $S \propto r_s^{D-2}$. It is also the reason that the size of quantum gravity corrections is set by $G_N/r_s^{D-2}$.    

The classical type IIA gravity action has two similarities. 
 The first corresponds to changing the string coupling and the RR field strengths 
 \be \la{nuSym}
 e^{ - 2 \phi } \to \nu^2 e^{-2 \phi } ~,~~~~~~~  F_q \to \nu F_q 
 ~,~~~~~~~g_{\mu \nu} \to g_{\mu \nu } ~,~~~~~~~B_{\mu \nu} \to B_{\mu \nu } ~,~~~~~~~S\to \nu^2 S 
 \ee 
 In our case, since $N$ is proportional to the flux of a RR field strength,  this similarity implies the familiar result that the action scales like $N^2$. 
 This similarity is   exact in $\alpha'$ and it extends to the weakly coupled regime (but not to the very strongly coupled regime, below the lower limit in \nref{Range}). In the matrix model it follows from large $N$ counting involving planar diagrams.  
 
 The second similarity of the type IIA supergravity action is 
 \be \la{TenRes}
 g_{\mu \nu} \to \Omega^2 g_{\mu \nu} ~,~~~~~~~ \phi \to \phi~,~~~~~~~ F_q \to \Omega^{ q-1 } F_q ~,~~~~~~~B_{\mu \nu} \to \Omega^2 B_{\mu \nu} ~,~~~~~~~S \to \Omega^8 S ~,
 \ee  
 which is simply the statement that we have a two derivative action. 
The rescaling \nref{ResCa} does not leave the metric or the gauge and scalar fields invariant. Instead, it changes them  in the same way as a particular combination of the similarities \nref{nuSym} and \nref{TenRes}, with  
 \be 
 \Omega = \gamma^{ 3/10} ~ ,  ~~~~~\nu = \gamma^{-21/10} ~,~~~~~\Omega^7 \nu =1
 \ee 
 where the last relation ensures that $N$ is not changed under \nref{ResCa}.  This implies that the action changes as in \nref{ActExp}.

 The scaling similarity \nref{ResCa} implies that the action  and the entropy of the finite temperature solution behave as 
 \be 
 S \propto N^2 T^{9/5} \propto  N^2 T^{9/5} \lambda^{ -3/5} \la{TempEnt}
 \ee 
 where in the last step we restored the $\lambda $ dependence by dimensional analysis. Here we used that the temperature changes as $T \to \gamma^{-1} T$ under \nref{ResCa} since it is proportional to $1/\beta$,  and $\beta$ is the length of Euclidean time which changes as $\beta \to \gamma \beta$ according to \nref{ResCa}. Note that the $N^2$ factor is fixed by \nref{nuSym}. 
 
 This similarity of the asymptotic form of the metric can also be used to classify the various perturbations of the theory. They can be characterized by the decay of normalizable perturbations  at large  $\rho$, or small $z$, 
 \be 
  \chi   \propto z^{\Delta } ~,~~~~~{\rm as}~~~~ z\to 0 
  \ee 
 % where $\chi$ is some gravitational perturbation of the solution. 
 We can then think of  $\Delta$ as the dimension of the corresponding operator. We will explain this in more detail later.

 Let us stress that neither $N$ nor $\lambda$ are changed when we apply the transformation \nref{ResCa}.  %Note that under the rescaling similarity \nref{ResCa} we are keeping $N$ and $\lambda $ fixed. 
A similar sounding scaling symmetry was discussed in \cite{Sekino:1999av,Jevicki:1998ub}\footnote{They also considered in a addition a special conformal symmetry of a similar kind, with a time dependent change in the coupling.}  where  they changed $\lambda \to \gamma^{-3} \lambda$ (keeping $N$ fixed). This is an exact ``symmetry'' of the full matrix model which is usually called dimensional analysis.  We put ``symmetry'' in quotation marks because it involves changing a coupling constant.  We emphasize that the scaling similarity \nref{ResCa} is {\it not} dimensional analysis. It is a non-trivial similarity that emerges at low enough energies and reflects a non-trivial critical behavior of the matrix model. In particular, \nref{ResCa} is definitely not a symmetry (or a similarity\footnote{The concept of similarity depends on a notion of classical limit. What we are saying is that \nref{ResCa} is not   a similarity of the classical matrix model action.  The matrix model action does have a similarity, which is that of a usual quartic potential in non-relativistic classical mechanics and it fixes the very high temperature dependence obtained in \cite{Kawahara:2007ib}.}) of the classical  matrix model action. 

 This rescaling similarity is somewhat analogous to the conformal symmetry of the SYK model.  Both emerge at relatively low temperatures.   The parameter $J$ in SYK is analogous to $\lambda^{1/3}$ here. Both set the scale at which the model becomes strongly coupled. And in both cases, the critical behavior is modified when we go to temperatures that are parametrically small in the $1/N$ expansion. 
 
 The rescaling similarity is {\it not} the symmetry associated to      the 11 dimensional boost symmetry that should emerge at extremely low energies  according to the BFSS conjecture \cite{Banks:1996vh}. That is yet another emergent symmetry appearing at lower energies\footnote{ For completeness, let us mention that this new symmetry necessary for the BFSS conjecture keeps $g^2$ fixed and $r$ fixed and sends $N\to \zeta N $, $t \to \zeta t$ (or $E \to \zeta^{-1} E $) \cite{Banks:1996vh}. For the thermal states, it should start appearing for termperatures lower than the Gregory Laflamme instability, or for ${T \over \lambda^{1/3}} \ll  N^{ - 5/9}$. }.     Nevertheless, the similarity does have a  some connection to   eleven dimensional  boosts   as we explain in section \ref{ElevenD}.

 \subsection{Eleven dimensional uplift}
 \la{ElevenD}

When  $e^\phi$ becomes large, the metric \nref{SolIIA} is no longer a good description. But we can use an eleven dimensional metric which is simply the Kaluza Klein uplift of \nref{SolIIA} given by 
%
 %The eleven dimensional lift of \nref{SolIIA} is
 \bea \la{KKGen}
  {d s^2_{11} \over l_p^2 } & = & e^{4  \phi/3 } \left(d \varphi - \frac{A}{\sqrt{\alpha'}}\right)^2 + e^{- 2  \phi/3 } { ds^2_{str} \over \alpha'} 
 \\
 &=& - 2 d \hat \tau d \varphi + {\alpha \over y^7} d\varphi^2 + {y_0^7 \over \alpha } d\hat\tau^2 +  { d y^2 \over h } + y^2 d \Omega_8^2   ~,~~~~~~~~h = 1 - { y_0^7 \over y^7}    \la{11dmet}
 \\ 
 \alpha &=& { d_0 N \over (2\pi)^2}  ~,~~~~~~~ \hat \tau = \tau \alpha^{-1/3} = t (2\pi g)^{2/3}  ~,~~~~~~y = \rho \alpha^{1/3}
 \eea  
 where $\varphi \sim \varphi + 2 \pi $.

 This is the metric of a plane wave in eleven dimensions. In this form of the metric the dimensionless time $\hat \tau$ is given in terms of the energy scale $ g^{2/3}$.  
   
 The range of validity of this eleven dimensional solution is \cite{Itzhaki:1998dd}
 \be \la{Range}
  N^{-5/9 } \ll { T \over \lambda^{1/3}} \ll N^{ -10/21} 
  \ee 
 The lower limit is due to the appearance of a Gregory Laflamme instability \cite{Gregory:1993vy}\footnote{This can be seen by rescaling the above metric $y = N^{1/9} y'$, $\varphi = \varphi'$ and  $\hat \tau = N^{2/9} \hat \tau'$. The metric then becomes proportional to   $ds^2 \propto - 2 d\hat \tau' d\varphi + { d\varphi^2 \over {y'}^7}  + d\vec {y'}^2 $. Then we expect that, once we go to finite temperature,  the GL instability appears for $\hat \beta' \sim 1$ which translates into the lower limit in \nref{Range}.}.  
   
Since the eleven dimensional metric is related to the ten dimensional metric by the simple transformation \nref{KKGen},  it inherits the rescaling similarity \nref{ResCa}, 
\be \la{ResSim11}
 \hat \tau \to \gamma \hat \tau ~,~~~~~~y\to \gamma^{-2/5} y ~,~~~~~~ N \to N ~,~~~~~g^2 \to g^2 ~,~~~~~\varphi \to \gamma^{-9/5} \varphi 
 \ee 
 The meaning of the rescaling of $\varphi$ requires some explanation because $\varphi$ is a periodic variable, and we will {\it not} be changing its period. To explain it, let us note that this set up, \nref{KKGen} also has the two similarities \nref{nuSym} and \nref{TenRes} present in general IIA theories. 
 
  The most obvious similarity of the eleven dimensional metric is a simple transformation in which the whole eleven dimensional metric is rescaled,
 \be \la{ResEl}
 g_{\mu \nu} \to \tilde \Omega^2 g_{\mu \nu} ~,~~~~~~~S \to \tilde \Omega^9 S \ee 
% where $g_{\mu \nu}$ is the eleven dimensional metric.
 We distinguish this from the one we had in ten dimensions \nref{TenRes}. 
 
 Now, a metric that is translation invariant along the $\varphi$ direction also has a similarity under replacing $\varphi \to \varphi /\kappa$ in the metric ansatz \nref{KKGen}. 
 What we mean is that we keep $\varphi$ periodic but we redefine  $\phi, ~A,$ and  $g_{str}$ so as to absorb $\kappa$.   
 More explicitly, the transformation is 
 \begin{align} \la{kappaDef} 
  e^{ 4 \phi \over 3 } ( d \varphi  - A)^2 + e^{ - { 2 \phi \over 3} } ds^2_{10, str}  & \ra     e^{ 4 \phi \over 3 } \left( {d \varphi \over \kappa}   - A\right)^2 + e^{ - { 2 \phi \over 3} } ds^2_{10, str} \\
 & =  e^{ 4 \phi' \over 3 } \left( {d \varphi }   - A'\right)^2 + e^{ - { 2 \phi' \over 3} } d{s'}^2_{10, str}   ~,~~~~~~~~~~ S \to \kappa^{-1}  S 
 \\
  ~ {\rm with }~~~~~~~~ A' = \kappa A ~,~~~~~~~~&e^{ \phi'} = \kappa^{-3/2} e^\phi ~,~~~~~~~~g'_{10,str, \mu \nu } = \kappa^{-1} g_{10,str,\mu \nu } 
\end{align}
   
%    \bea \la{kappaDef} 
%  e^{ 4 \phi \over 3 } ( d \varphi  - A)^2 + e^{ - { 2 \phi \over 3} } ds^2_{10, str}   \ra     e^{ 4 \phi \over 3 } \left( {d \varphi \over \kappa}   - A\right)^2 + e^{ - { 2 \phi \over 3} } ds^2_{10, str} \\
%  = &  e^{ 4 \phi' \over 3 } \left( {d \varphi }   - A'\right)^2 + e^{ - { 2 \phi' \over 3} } d{s'}^2_{10, str}   ~,~~~~~~~~~~ S \to \kappa^{-1}  S 
% \\
%  ~& {\rm with }~~~~ A' = \kappa A ~,~~~~~e^{ \phi'} = \kappa^{-3/2} e^\phi ~,~~~~~g'_{10,str, \mu \nu } = \kappa^{-1} g_{10,str,\mu \nu } 
%   \eea
% where we we view the second expression  as defining new fields $\phi, ~A,~g_{10,str}$ after we put the ansatz in the same form as in \nref{KKGen}\footnote{Just to be completely explicit, $A \to \kappa A$, $e^{\phi} \to e^{\phi} \kappa^{ -3/2}$, $g_{10, \mu \nu} \to \kappa^{-1}  g_{10, \mu \nu}$.}.  For example, the transformation for 
% $A$ is $A\to \kappa A$. We have also indicated how the action changes. 
The action would be invariant if we were to change the period of $\varphi$, but since we do not change it, the action changes as indicated. 
   This is obviously a symmetry of the equations of motion because it can be viewed as a simple coordinate transformation for $\varphi$, once we forget about its period.  
     
  Of course, each of the two transformations \nref{ResEl} \nref{kappaDef} corresponds to a combination of the two transformations \nref{nuSym} \nref{TenRes}. 
  
%  As in the IIA discussion, the rescaling similarity 
 % $t \to \gamma t $, $y \to \gamma^{-2/5} y$, is combined also with two similarity transformations of 11 dimensions generated by 
  The rescaling similarity \nref{ResCa}, which in the coordinates of \nref{11dmet} amounts to $\hat \tau \to \gamma \hat \tau, y \to \gamma^{-2/5} y$, %, \varphi \to \varphi$,
   does not leave the eleven dimensional metric invariant. Instead, it changes it in the same way as a particular combination of the two similarities  \nref{ResEl} \nref{kappaDef}  \be 
  \tilde \Omega = \gamma^{ -2/5} ~,~~~~~~~~\kappa = \gamma^{- 9/5}
  \ee 
 % 
  %\AB{I have understood the above sentence to mean, ``The IIA %rescaling similarity \nref{ResCa}, which in the coordinates of \nref{11dmet} amounts to $\hat \tau \to \gamma \hat \tau, y \to \gamma^{-2/5} y, \varphi \to \varphi$, implements a combination of the two M-theory similarities  \nref{ResEl} and \nref{kappaDef} with the parameters 
%  \begin{align}
%  	\tilde{\Omega} = \gamma^{-2/5}, \quad \quad \kappa = \gamma^{-9/5}
%  \end{align}
%Is this correct?  I was confused at first whether the coordinate rescaling that you referred to involved $\varphi$ or not. Maybe it would be clearer to write it in some way like this.}
    So the rescaling of $\varphi$ in \nref{ResSim11} should be interpreted as implementing the M-theory similarity transformation \nref{kappaDef}.

%It is also useful to emphasize that \nref{ResSim11} is a combination of a rescaling of all the coordinates by $\gamma^{-2/5}$  combined with a boost $b$ that acts as\footnote{Note that the transformations suggest that we should think of $\tau = x^+$ and $\varphi = x^-$ where $x^\pm$ are the usual light cone variables. Here $x^-$ is compact \cite{Susskind:1997cw}. } 
%\be \la{EBoost}
%b: ~~~~~~~~~ \hat \tau  \to \gamma^{7/5} \hat \tau ~,~~~~~~~~ \varphi \to \gamma^{ - 7/5} \varphi ~,~~~~~~~\vec y \to \vec y 
% \ee 

 This eleven dimensional picture is particularly useful for determining the dimensions of the operators. The reason is that we can look at the behavior of the fields at large $y$, where they are simply perturbations around flat space. We can then expand the fields in powers of $y$ and look at their eigenvalue with respect to the transformation \nref{ResSim11}. For this analysis, we note that \nref{ResSim11} is a combination of a rescaling of all the coordinates 
\begin{align}\label{Res}
	\hat \tau \to \gamma^{-2/5} \hat \tau ~, \quad \quad \varphi \to \gamma^{-2/5} \varphi~, \quad \quad \vec{y} \to \gamma^{-2/5} \vec{y}
\end{align}
with a boost $b$ that acts as\footnote{Note that the transformations suggest that we should think of $\tau = x^+$ and $\varphi = x^-$ where $x^\pm$ are the usual light cone variables. Here $x^-$ is compact \cite{Susskind:1997cw}. } 
\be \la{EBoost}
b: ~~~~~~~~~ \hat \tau  \to \gamma^{7/5} \hat \tau ~,~~~~~~~~ \varphi \to \gamma^{ - 7/5} \varphi ~,~~~~~~~\vec y \to \vec y 
 \ee

 At large $y$ the perturbations can be expanded as $\chi \sim y^\ell $, where $y^\ell$ denotes a symmetrized traceless homogeneous degree $\ell$ polynomial (it is traceless because it needs to obey the flat space Laplace equation).  The rescaling of $y$ in \nref{Res} will contribute to the transformation of the field. Let us consider,  for example, a perturbation of the metric component $\delta g_{ij} \sim y^\ell $ along the transverse dimensions. This will scale like $\gamma^{ - 2 \ell/5}$. When we compute the scaling of $\delta g_{ij}$, we pull out an  overall factor of $\gamma^{ - 4/5}$ that comes from rescaling $dy^i dy^j$ in $\delta g_{ij} dy^i dy^j$. 
This is because the dimensions of the fields come from the scaling of the metric fluctuations relative to the scaling of the background. 
From this scaling we can read off the dimension as we explain below.

 We are considering fields that are constant in time and  growing at infinity. They can be viewed as the result of adding an operator to the action 
 \be \la{ActPer}
  S_{pert} = \zeta  \int dt O
 \ee 
 In a conformal theory the growth of the field in the bulk is related to dimensions of $\zeta$. 
 In other words, the operator insertion corresponds to a  bulk  field with a non-normalizable component going as 
  \be \chi \to  \zeta z^s ~~~~~~~~{\rm as~~~ } ~~~ z\to 0 
  %\gamma^{s} \chi ~,~~~~~~({\rm because } ~~\chi \propto z^s)
 \ee 
 %under rescaling, 
 We see that if we assign  $\zeta$ a  dimension $s$ then we can keep $\chi$ invariant under scaling.   In a $d$ dimensional  CFT, the analog of $S_{pert}$ in \nref{ActPer} should be invariant. This implies that 
 \be 
 s = d-\Delta  
 \ee 
  where $\Delta$ is the dimension of $O$. 
 
 In our case, the similarity rescales the action. Therefore its natural to require that $S_{pert} $ scales as $S_{pert} \to \gamma^{-\etanew   } S_{pert}$ with $\etanew   = 9/5$. 
 This means then that we should assign to $O$ the dimension $\Delta$ such that 
 \be \la{DimPert}
 s = 1 + \etanew   -\Delta  =    d -\Delta ~,~~~~~~~~~d \equiv 1 + \etanew = { 14\over 5} 
 \ee 
  where we defined $d$   in analogy with formulas we have in a CFT. In section \ref{FUplift} we will see another sense in which $d$ analogous the dimension of the $AdS$ boundary. 
  
  This means that the $\delta g_{ij}$ metric components that go like $y^\ell$ have $s= - 2\ell/5$, which, using \nref{DimPert}, leads to   
  \be 
  \Delta = d + { 2 \ell \over 5} = { 14 \over 7 } + { 2 \ell \over 5} 
  \ee 
  
 In cases that the metric fluctuation has $\hat \tau$ or $\varphi$ indices, then 
there is another contribution to the scaling that comes from the boost eigenvalue of the fields. As noted above, the scaling similarity \nref{ResSim11} can be viewed as a combination of \nref{Res} plus the boost \nref{EBoost}. The boost leads to extra factors of $\gamma$ 
%When we perform the scaling similarity \nref{ResSim11}, we are also performing the boost \nref{EBoost}, which leads to extra factors of $\gamma $ 
depending on 
 how many $\hat \tau$ or $\varphi$ indices the field has.  We find that the possible eigenvalues under \nref{ResSim11} are then 
 \be \la{OpDim}
s= d - \Delta  = -b { 7 \over 5 }  - { 2 \over 5 } \ell ~,~~~~~~ \Delta = ( 2 + b ) { 7 \over 5 } + { 2 \over 5 } \ell  ~,~~~~~~~~~b =-2,-1,0,1,2~,~~~~~~d \equiv { 14 \over 5 } 
 \ee 
 where $b$ is the boost eigenvalue of the field.   For example $\delta g_{\hat \tau , i}$ has $b=-1$.  It might sound a little strange that the boost eigenvalue contributes to the scaling dimension, since the latter might seem to involve only the powers of $y$.
 To explain this more clearly, it is convenient to write both the background metric \nref{11dmet} and its fluctuations in terms of vielbeins as follows (for $y_0=0$) 
 \bea 
 ds^2 &=& - 2 e^{\hat \tau } e^{\hat \varphi} + \alpha e^{\hat \varphi} e^{\hat \varphi} + e^{\hat i } e^{\hat i} + \delta h_{\hat \mu \hat \nu } e^{\hat \mu } e^{\hat \nu} \la{MetFlVb}
 \\
 &~&   e^{\hat {\hat \tau }} =y^{7/2} d\hat \tau ~,~~~~~~e^{\hat \varphi } = y^{-7/2} d\varphi ~,~~~~~~~ e^{\hat i} = dy^i 
 \eea

% \AB{It seems like the fix did not sync...I am not sure why. It  seems to me that the vielbein is missing a couple things. One is a term that would give the $\frac{y_{0}^{2}}{\alpha}d\hat \tau^{2}$ in the original background metric. The other is that if are writing the vielbein for the background before taking the large $y$ limit then it seems we should also account for the factor of $h^{-1}$ in the $dy^{2}$ component. I was trying to work out what the vielbein is that does this and was getting something complicated...I will try this again later}
 All the vielbeins transform as $ e^{\hat \mu} \to \gamma^{-2/5} e^{\hat \mu } $ under \nref{ResSim11}. Here the $\hat \mu$ and $\hat \nu$ indices run over all eleven values. 
 It is natural to think that the canonically normalized fields will be the metric fluctuations $\delta h_{\hat \mu \hat \nu}$ that multiply the vielbeins as in \nref{MetFlVb}. Compared to the naively defined metric flucutations $\delta h_{\mu \nu}$,
 %, for example $\delta g_{\varphi \varphi } d\varphi^2 $, 
 they have extra powers of $y^{7/2}$ which depend on how many $\varphi$ or $\hat \tau$ indices it has. More precisely, they depend on the boost eigenvalue of the field. So, for example a fluctuation of $\delta g_{\varphi \varphi } \sim y^\ell $ leads to a 
 $\delta h_{\hat \varphi \hat \varphi } \sim y^{ \ell + 7} $ and a transformation under scalings which leads to \nref{OpDim} with $b=2$. 
% %one option would be to instead use a tilde
%  To make this more plausible, consider the metric fluctuation $\delta g_{\varphi \varphi } d\varphi^2$. We are saying that if $\delta g_{\varphi \varphi} \sim y^\ell$, then the total scaling, including the boost, is $\gamma^{ -14/7 - 2 \ell/5}$. This is reasonable if we think that we should compare 
% $\delta g_{\varphi \varphi} $ to the original metric that contains a term ${ d\varphi^2  \over y^7}$, so that the ratio between the two is $y^7 \delta g_{\varphi \varphi}$. Then final $y$ dependence leads to $z^s$ with $s=- d - { 2  \ell /5}$ or a total $\Delta = { 28\over 5} + { 2 \ell \over 5 }$.  In general, we expect that in order to transform the raw metric fluctuation into a normalized field on the background geometry we will need factors of $y$ related to the boost eigenvalues, so that the final scaling is indeed what we read off from \nref{OpDim}. 

  The fermionic fields have half integer values of $b$. We have also defined $\Delta$, whose precise meaning is explained below. 
   A full table of operators with the precise lower bounds on the possible values of $\ell$ for each case is given in \cite{Sekino:1999av}\footnote{$\nu_{\rm theirs} = \Delta_{\rm ours} -d/2 = \Delta_{\rm ours} -7/5$ and $\Delta_{\rm theirs}$ is not equal to what we called $\Delta$ here.}.

  The set of dimensions computed in  \nref{OpDim} is the same as what we would obtain if instead we looked at the decaying part of the field, where we define the dimension via $\chi \propto z^\Delta $. In this case, we look at fields in flat space going like 
    $y^{-7 - \ell}$.  This also gives the set \nref{OpDim} after we take into account the boost eigenvalues of the fields.

We presented here a quick way to read off the dimensions. The answers agree with the detailed analysis discussed in  \cite{Sekino:1999av}.

The bosonic supergravity modes with dimensions given in \nref{OpDim} transform under the following irreducible representations of $SO(9)$. For $\ell\geq 2$ the metric gives rise to 
  a tensor mode transforming under the $SO(9)$ representation labelled by the highest weight vector $(\ell, 2, 0, 0)$, two vector  modes transforming under the $(\ell, 1, 0, 0)$ representation, and three scalar modes transforming under the $(\ell, 0,0,0)$ representation \cite{Sekino:1999av,Rubin}. These representations are labelled by the Young diagrams \cite{gourdin1982basics}

\begin{center}
$
\overbrace{~\ydiagram{3} ~ . \quad . \quad . ~\ydiagram{1}^{~T}}^{\ell} ~,~~~\overbrace{~\ydiagram{3,1} ~ . \quad . \quad . ~\ydiagram{1}^{~T}}^{\ell} ~,~~~~~\overbrace{~\ydiagram{3,2} ~ . \quad . \quad . ~\ydiagram{1}^{~T}}^{\ell}
$
\end{center}
 
These modes arise from the fields $\delta g_{ij}$, $\delta g_{\hat \tau i}$, $\delta g_{\varphi i}$, $\delta g_{\hat \tau \varphi} $, $\delta g_{\hat \tau \hat \tau}$, $\delta g_{\varphi \varphi}$ and their dimensions are given by \nref{OpDim} where the boost eigenvalue depends on the number of $\hat \tau$ and $\varphi$ indices\footnote{ For $\ell =1$ we have just one vector and two scalars. For $\ell =0$ we only have one scalar. The scalar  representations we lose for these lower values of $\ell$ are those with the lowest conformal dimension according to \nref{OpDim},  among the three   scalar  representations for that value of $\ell$. Similarly, the missing $\ell =1$ vector representation is the one with the lower $\Delta$ of the two towers \cite{Sekino:1999av}. }.
  Similarly the three form $C_{\mu \nu \delta}$ leads to one $SO(9)$ vector representations labelled by the highest weight vector $(\ell, 1, 0, 0)$, two antisymmetric tensors  transforming under the $(\ell, 1, 1, 0)$ representation, and an antisymmetric tensor in the   $(\ell, 1,1,1)$ representation. We get one for each $\ell\geq 1 $ \cite{Sekino:1999av}.  These have the following Young diagrams 
 \begin{center}
$
\overbrace{~\ydiagram{3,1} ~ . \quad . \quad . ~\ydiagram{1}^{~T}}^{\ell}
~,~~~~\overbrace{~\ydiagram{3,1,1} ~ . \quad . \quad . ~\ydiagram{1}^{~T}}^{\ell}
 ~,~~~~~~~
\overbrace{~\ydiagram{3,1,1,1} ~ . \quad . \quad . ~\ydiagram{1}^{~T}}^{\ell}
$
\end{center}  
These come from the $C_{\hat \tau \varphi i } $, $C_{\hat \tau ij}$, $C_{\varphi ij}$ and $C_{ijk}$ components and the dimensions are again given by \nref{OpDim}. 
 \subsection{Matrix model computation of the critical exponent $\etanew  $} 

 \la{MMDer}

 From the matrix model point of view the scaling similarity is an emergent similarity. In other words,  it is not obvious from  the matrix model side. An important property of the similarity is the action scaling exponent \nref{ActExp}. In this section, we discuss a computation of this scaling exponent assuming that we have a similarity with an unknown action scaling exponent $\hat\theta$,   and then computing the exponent using  the constraints of supersymmetry.    
 
 We start from the matrix model and assume that we have an emergent similarity transformation $t\to \gamma t$. Then we Higgs the gauge group to $U(N/2) \times U(N/2)$ by giving a diagonal vev to one of the nine  bosonic matrices of the form $X^1 = x \sigma_3 \otimes {\bf 1}_{N/2}$.   This vacuum expectation value spontaneously  breaks the scaling similarity. Since this similarity is a symmetry of the action, it should also be a symmetry of the effective action for this expectation value. To facilitate the analysis we define  a new coordinate $z$ which is a yet to be determined power of $x$, $z =x^a$ for some power $a$. $z$ is chosen so that  it  transforms as $z \to \gamma z$ under the scaling similarity. 
   
 We now imagine that $z$ (or $x$) is changing slowly. The effective action for this motion is constrained by the scaling similarity to be of the form 
  \be 
  S \sim  N^2 \int dt z^{- \hat \theta   -1} f( \dot z^2 ) ~,~~~~~~~f = f_0 + f_2 \dot z^2 + f_4 \dot z^4 + \cdots 
  \ee 
  where $f_i$ are some constants and the overall factor of $N^2$ comes from the usual 't Hooft scaling. We write a hat in $\hat \theta   $ because it is still an unknown number, the action scaling exponent.  Of course we want to argue that $\hat \theta   = \etanew  $ in the end. 
  Supersymmetry puts some constraints on these coefficients. First, supersymmetry   implies  that $f_0=0$. 
  Then it says that the second term, the kinetic term, is not renormalized if we express it in terms of the diagonal components of the matrix 
  \be \la{Xandz}
   \dot x^2 =  f_2 z^{-\tilde d} \dot z^2  \longrightarrow  x \propto  z^{- ( \tilde d-2) \over 2} ~,~~~~~\tilde d = \hat \theta  +1
 \ee 
 This relates the coordinate $x$ to the coordinate $z$. 
 Then the quartic term in velocities becomes 
 \be 
 f_4 z^{-\tilde d} \dot z^4 \propto   \dot x^4  x^{ - {2 \tilde d \over \tilde d -2} } 
 \ee   
It was argued using supersymmetry that the power of $x$ should be $-7$ \cite{Paban:1998ea}  and, furthermore, the coefficient is one loop exact\footnote{In fact, \cite{Paban:1998qy} showed that also the $v^6$ term is protected, though we do not need it for this argument. }. 
 This then gives the equation 
 \be \la{DandEta}
 7 = { 2 \tilde d \over \tilde d -2} ~,~~~~~\to ~~~~~\tilde d = { 14 \over 5} =d ~,~~~~\hat \theta   = \etanew   = { 9 \over 5 } 
 \ee 
 
 This can be then viewed as a derivation, from the matrix model, of the action scaling exponent for the similarity transformation. This scaling exponent determines the temperature dependence of the action as we saw in \nref{TempEnt}. 
 
 This computation also implies that the low energy action on the moduli space  
 \be \la{vfourth}
 S \propto N^2 \int dt \left[  \dot x^2 + \# { \dot x^ 4 \over x^7 } \right] 
 \ee 
 also has a scaling similarity with the same exponent $\etanew  $. This fact underlies the observation in \cite{Smilga:2008bt,Wiseman:2013cda}, who   proposed an explanation for the entropy of these black holes using the moduli space approximation\footnote{Here we are saying that their explanation boils down to the observation that the two systems have a scaling similarity with the same exponent. Whether there is a precise relation or not, we are not sure. Here we explained why the exponent had to be this one.  }. 
  
  From the gravity point of view,  the full action for a brane probe $S \propto \int dt z^{-d} [ \sqrt{ 1 - \dot z^2} -1] $ also has the scaling similarity, with $d$ as in 
  \nref{DandEta}\footnote{ See also \cite{Jevicki:1998ub} for further discussion on symmetry constraints on this action.}.  
  
  A curiosity is that the dimension of $x$ in \nref{Xandz} is such that it scales as a free field in $d$ dimensions. We should also mention that the non-renormalization of the $\dot {\vec x}^2$ term, including the angular components, implies that the ratio of the AdS and sphere radii should be as obtained from the gravity solution \nref{AdSCo}.

   We can also use this non-renormalization argument to fix at least some of the operator dimensions. 
   %We can imagine giving a vev to the diagonal components of the form 
   %$\vec x_i$ to $N-k$ and the same one $\vec x$ to $k$ of them. \AB{Here is a rephrasing of the previous sentence that I think is clearer: ``
   We can imagine giving vevs $\vec{x_{i}}$ to $N-k$ of the diagonal components and the same vev $\vec{x}$ to the other $k$ components. If $k\ll N$ we expect that we can approximate the solution in terms of a brane probe on a background geometry.   The brane probe action has terms of the form 
   \be \la{ProbeMa}
   \int dt  (\dot {\vec x})^4 \sum_i { 1 \over |\vec x - \vec x_i |^7 } \propto \sum_\ell \int dt (\dot {\vec x})^4 \sum_{i  } { (\vec x_i)^\ell (\hat x)^\ell \over |x|^{ 7 + \ell } } \sim  \sum_\ell  \int dt (\dot {\vec x})^4 { \hat x^{a_1} \cdots \hat x^{a_\ell}  \over |x|^{7+\ell}} O_{ a_1 \cdots a_\ell} 
    \ee 
    where the last term is just an expansion in scalar spherical harmonics. We also defined $\hat x= { \vec x /|x|}$.  Supersymmetry determines the form of the first expression for the action, which implies that the action has a rescaling similarity. The second term is the schematic form of the expansion for $|\vec x| \gg |\vec x_i|$.  In the last term we summarized the average of $x_i^\ell $ as the expectation value of an operator, $O_{a_1 \cdots a_\ell}$,  in the theory of the $N-k$ branes.  Imposing that the first and last terms in \nref{ProbeMa} transform in the same way under the scaling similarity implies that   the dimension of the operator is 
    \be 
    \Delta =  { 2 \over 5 } \ell 
    \ee 
   We can similarly compute the dimensions of other operators by looking at terms involving velocities of the $N-k$ branes and so on. 

Let us emphasize that the form of the low energy action \nref{ProbeMa}  is constrained by supersymmetry even in the region where we do not have the scaling similarity, for example in the weakly coupled region, $\lambda/x^3 \ll 1$. 
%So that we needed to make the non-trivial assumption that we had a similarity in order to fix the action scaling exponent $\etanew $. Of course, it is also desirable to have a first principles matrix model derivation for the emergence of the similarity itself.  Here we only derived the exponent $\etanew$ once we assume that a similarity (with an unknown exponent) is present. \JM{Added this paragraph. Is it too repetitive?} \AB{No I don't think it is too repetitive, except for possibly the last sentence. The last 3 sentences of the paragraph could be condensed to: ``
So we needed to make the non-trivial assumption that we had a similarity (with an unknown action scaling exponent) in order to fix $\etanew $. Of course, it is also desirable to have a first principles matrix model derivation for the emergence of the similarity itself.
%''}

%\JM{I added the following paragraph, prompted by a question from R. Monten} 

  The weakly coupled theory also has a scaling similarity, which is the similarlity of a quartic classical action of the schematic form $I = \int dt Tr[\dot X^2 + [ X, X]^2 ] $, with $t \to \gamma t, ~ X \to \gamma^{-1} X, ~I \to \gamma^{-3} I$. On the other hand, the action  \nref{vfourth} is also valid in the weakly coupled region, $\lambda/x^3 \ll 1$. However, \nref{vfourth} does  {\it not} have that similarity of the weakly coupled action. In the weakly coupled  region, the $\dot x^4/x^7$ term in \nref{vfourth} should be viewed as a quantum correction which breaks the similarity of the classical theory. 
  On the other hand, in the strong coupling regime we were assuming that we have a scaling similarity, whose $\hbar$ is given by $1/N^2$. Therefore, in that regime it is reasonable to expect that both terms in \nref{vfourth} should be compatible with the similarity, since they are both part of the classical  action when the $\hbar$ expansion is the $1/N^2$ expansion\footnote{We thank R. Monten for a question leading to this paragraph.}.

 \subsection{$AdS_{2 + \etanew }$ uplift as a mathematical trick}
 \la{FUplift}
  
 In this section, we discuss a mathematical trick which leads to a simple way to derive the wave equation for fluctuations around the near extremal geometry \cite{Kanitscheider:2009as}. In addition, we will get another perspective on the similarity transformation by relating it to a more familiar $AdS$ situation. We do not claim that this trick has any physical meaning, it is just a mathematical trick.  
 
 The trick involves viewing the ten dimensional action involving the metric, dilaton, and gauge field as coming from a higher dimensional action involving only the metric and the gauge field, reinterpreting the dilaton as the volume of the extra dimensions. It is convenient to dualize the field strength in ten dimensions and view it as an eight form, $F_8 = *_{10} F_2$, whose flux on the 8-sphere is $N$, 
 \be 
  { 1 \over 16 \pi G_N } \int_{S^8} F_8  = { 1 \over 16 \pi G_N} \int_{S_8} * F_2 =  {  N \over \sqrt{\alpha'}}
 \ee 

  We now go to $10+\etanew $ dimensions, keeping the 8-form an 8-form in the higher dimensional space. We will assume that this metric has the form 
 \be \la{UpLi}
 d\hat s^2 = e^{a \sigma } d  s^2_{10} + e^{ 2 \sigma } d{\vec x}_\etanew  ^2 
 \ee 
 where $d {\vec x}_\etanew  ^2$ is the flat metric in $\etanew  $ dimensions and  $d s^2_{10}$ is the ten dimensional string metric.  Dimensionally reducing to ten dimensions we obtain the original string frame action (see appendix \ref{Reduction}) after choosing
 \be \la{etaphi}
 \etanew   = 9/5 ~,~~~~~~a ={ 3 \over 5} ~,~~~~~~~ \phi = - { 21 \over 10 } \sigma 
 \ee 
 The solution becomes very simple in the higher dimensional space, it is just $AdS_{2+\etanew  }\times S^8$, 
 where the ratio of the two radii is, as in \nref{AdSCo}, 
 \be 
  { R_{AdS_{2+\etanew  } } \over R_{S^8} }  = { 2 \over 5 } \la{RatRadii}
  \ee 
  
  When we deal with this higher dimensional metric, we will take all metric components independent of the extra dimensions and we will only allow a metric perturbation by the field $\sigma$ \nref{UpLi} in the $\etanew $ extra dimensions. In particular, there will be no metric fluctuations with indices in the extra dimensions.

  The near extremal solution is just the usual black brane  
  \be \label{bbmet}
   ds^2 \propto { 1 \over z^2 } \left[ - h d\tau^2 + dz^2/h + dx^2_\etanew   \right] ~,~~~~~~~~h = 1 - {z^d \over z_0^d} ~,~~~~~d = 1+ \etanew   = { 14 \over 5 } 
  \ee 
 % \AB{I think that in order to be consistent with the definitions of $t$ and $z$ in \nref{SolIIA} and \nref{AdSCo}, the time coordinate in the above solution should be $\tau$ instead of $t$:
 %   \be
 %  ds^2 \propto { 1 \over z^2 } \left[ - h d\tau^2 + dz^2/h + dx^2_\etanew   \right]  \ee 
 % }
  Under this uplift, the operators of dimension $\Delta$ are related to scalar fields with mass given by 
  \be \la{MassDel}
  m^2 R_{AdS}^2 = \Delta (\Delta -d) 
  \ee 
  where the set of dimensions is given by \nref{OpDim} in our case. This also implies that the two point functions of operators in the original ten dimensional solution are given by 
  \be \la{CorrSugra}
  \langle O(t) O(0) \rangle \propto { 1 \over |t|^{ 2\Delta -\etanew   } } ~,~~~~~~\etanew   = { 9 \over 5} 
  \ee 
  since they have the form of correlators in $d$ dimensions that have been integrated over $\etanew  $ of the dimensions in order to make them translation invariant along those $\etanew  $ dimensions. 
  This agrees with the expressions found in \cite{Sekino:1999av}. 
  
 \subsubsection{Uplifting $H_3$ and $F_4$} 
 
 The above discussion shows how to uplift the metric, dilaton and RR two form (dualized to an eightform). Here we just mention that we can also uplift the other forms in a rather formal way. 
 We need to take 
 \be 
 H_{3} \to H_{3 -\etanew/3} ~,~~~~~~~F_4 \to  F_{4+ { 2 \etanew \over 3} } 
 \ee 
where the subscripts indicate the type of form we have in the higher dimensional space. This simply reflects how the volume appears in the kinetic term of these forms.

   \subsection{Massive string states}
\la{MassStr}

Here we comment on the effective equation for massive string states. In this case, we expect an effective action of the form 
\be 
S \sim \int d^{10}x \sqrt{g}e^{ - 2\phi} \left[ (\nabla \chi)^2 + m^2 \chi^2 \right] \la{ActStt}
\ee 
where the metric is in the string frame, as in \nref{SolIIA}\footnote{Note that the mass term would not be a constant if expressed in the Einstein frame metric.}. This implies that in the semiclassical approximation of large mass we get an action of the form 
\be 
S = - m \int d\tau \sqrt{ g_{\mu \nu} \dot x^\mu \dot x^\nu } 
\ee   where $g_{\mu \nu}$ is 
  the string frame metric in \nref{SolIIA} and $m$ is the flat space mass of the string state ($m^2 = { 4 N/\alpha'}$ with integer $N$).
We will now consider the case with zero temperature, or $\rho_0=0$. 
We can write this action in terms of the $z$ variables in \nref{AdSCo} to obtain 
\be \la{Acz}
 S = - \left( {5 \over 2 } \right)^{3/10} { 2 \over 5}  \sqrt{\alpha'} m    \int { d\tau \over z }  z^{3/10 } \sqrt{ 1 -\dot z^2 } ~,~~~~~~\tau = ( d_0 \lambda)^{1/3}  t
 \ee 

Let us comment that from the point of view of the  higher dimensional $AdS_{2+\etanew  }$, this corresponds to the action of a massive particle with an effective mass  
\be \la{massz}
 R_{AdS}^2  m^2_{eff} = \alpha' m^2 z^{3/5} \left( {5 \over 2 } \right)^{3/5} {4 \over 25} 
 \ee 
 %where  
 % The $z$ dependence arises due to the overall $z$ dependence of the metric in \nref{AdSCo}. So the field is effectively light for small $z$ where the effective matrix model coupling is becoming small and it gets progressively heavier as we go to large $z$, where the effective  coupling of the matrix model is larger. 
 
 The two point function of the operators that insert this string state does not display a power law behavior at long Euclidean times.   Instead, the scaling similarity of the action \nref{Acz} implies that the two point function goes as   \be \la{CorrMS}
 \langle O(t) O(0) \rangle \propto \exp\left[  -C \sqrt{\alpha'} m \,( \lambda^{1/3} t ) ^{3/10 } \right] 
 \ee 
 in the geodesic approximation, where $C$ is a numerical constant that we can find by solving the classical problem associated to \nref{Acz}
 So we get an exponential decay rather than the power law decay we had for supergravity fields \nref{CorrSugra}.

 \subsection{Comments on relevant operators } 
 
   In this section we make some comments on the operator spectrum of the model in the scaling regime \nref{TemRa}.   
 In the UV theory, or the weakly coupled matrix model, the operator $X$ has dimension $-1/2$, so any operator of the form $Tr[X^\ell] $ is relevant. However, as we go to the infrared and the coupling becomes strong, many of these operators acquire high anomalous dimensions. If the operator is dual to a massive string state this dimension grows with scale as in \nref{CorrMS}. For the smaller subset of operators that correspond to gravity modes, the dimensions are given in \nref{OpDim}. 
 If one is interested in a quantum or classical simulation of this matrix model, an 
 interesting question is whether any of these operators are relevant. 
 First one can wonder what we mean by a   ``relevant'' operator. A relevant operator is one with  
 $\Delta < d$, $d=14/5$. This is the traditional condition for a relevant deformation in the $AdS_{2+\etanew}$ description. This is the correct condition since such an operator would give a growing deformation of the geometry, relative to the original geometry, as we go to the IR. 
 It is also the condition that implies that the coefficient of the perturbation in \nref{ActPer} has positive mass dimension according to \nref{DimPert}. 
 
 We see that there are a few single trace operators in \nref{OpDim} which are relevant. 
 \bea  \la{LowOp}
 Tr[ X^\ell ] & ~&  ~,~~~~~~ \Delta = { 2 \over 5 } \ell ~,~~~~~~ \ell = 2,3,4,5,6
 \cr 
 Tr[\dot X^i X^\ell ]  & ~&  ~,~~~~~~ \Delta = { 7 \over 5} + { 2 \over 5 } \ell ~,~~~~~~ \ell = 2,3
 \cr 
 Tr[ [X^i,X^j] X^\ell ]  & ~&  ~,~~~~~~ \Delta = { 7 \over 5} + { 2 \over 5 } \ell ~,~~~~~~ \ell = 1,2,3
\eea
where the indices are schematic \cite{Sekino:1999av}. For example, $Tr[X^\ell]$ really means $Tr[ X^{(a_1} \cdots X^{a_\ell )} ] $ where the indices are symmetrized and the traces are removed. We have also ignored fermion terms. For example the $\ell=1$ case of the last operator in \nref{LowOp} also contains a term 
\be 
\la{FerMass} Tr[\bar \psi \gamma^{ijk} \psi]
\ee 
 A particular component of this operator is turned on by the mass deformation discussed in \cite{Berenstein:2002jq}\footnote{ Notice that if we deform the action by this operator the coefficient $\zeta$ as in \nref{ActPer} would have dimensions of mass or energy, and is the $\mu$ parameter in \cite{Berenstein:2002jq}.}. 
 
The matrix theory operators corresponding to various supergravity interactions and currents were identified in \cite{Kabat:1997sa, Taylor:1998tv}; see there for a complete expression of the operators \nref{LowOp}. 

The significance of these operators is that they are the operators whose coefficients need to be fine tuned in order to get to the IR theory. The good news is that there are a small number of them. The bad news is that this number is nonzero, though this is not surprising when we want to get a quantum critical point.  Notice that none of these operators is an $SO(9)$ singlet. So if we could preserve the $SO(9)$ symmetry, then there are no single trace relevant deformations. Since we have only a finite number, it is perhaps possible that a discrete subgroup of $SO(9)$ is enough to remove all single trace operators, but we did not investigate this in detail. 
On the other hand, there are double and triple trace relevant operators that are $SO(9)$ singlets which are obtained by taking products of the operators in \nref{LowOp}. 
One could also have products of fermionic operators that we have not listed explicitly here. 

In addition to the $SU(N)$ singlet operators we discussed, in principle one could consider deformtations of the theory which are not $SU(N)$ singlets. It was argued in \cite{Maldacena:2018vsr} that $SU(N)$ non-singlet states have higher energies and excite string states with endpoints near the boundary of the gravity region. This suggests that it is likely that these deformations are also irrelevant. However, this should be more carefully analyzed.

  \section{Quasinormal modes } 
  
  \subsection{Some generalities about QNM}
  
  Quasinormal modes (QNM) characterize the response of the black hole to simple perturbations. They tell us how such perturbations decay at late times.  Here, late means large compared to $\beta$ but not compared to $N^2\beta$.   In the matrix theory, they describe how quickly perturbations of the thermal state relax back to equilibrium.
  In general we expect the QNM frequencies to be of order the temperature. In fact, the scaling similarity constrains them to be simple numbers times the temperature. This is because it implies that (\ref{SolIIA}) is a solution to the vacuum Einstein equations for any $\rho_{0}$. Then under  (\ref{ResCa}), solutions to the equations of motion in one background are mapped to solutions in the rescaled background: $e^{-i \omega(z_{0})t} =e^{-i \omega(\gamma z_{0})\gamma t}$. So $\omega \propto \frac{1}{z_{0}} \propto T$. We define
  \be 
   \omega_n =   2 \pi T \alpha_n   
   \ee 
   where $\alpha_n$ are dimensionless complex numbers. $T  = \frac{d}{4 \pi z_{0}}$ is the temperature with respect to $t$ in (\ref{bbmet}). The $\alpha_{n}$ can be computed by solving the wave equation with ingoing boundary conditions at the future horizon. In our case we also put Dirichlet boundary conditions at infinity, since the black hole is effectively in a box. 
   
 % Note that even though we have a large $N$ many body system, the set of modes is $N$ independent. 
  % in the sense that their frequencies are independent of $N$ Other simple perturbations decay faster. 
  
  {\bf Note added} 
 
 After submission, we learned that the first two QNM frequencies of the SO(9) scalar perturbation were first computed in \cite{Craps:2016cgo}\footnote{We thank K\'evin Nguyen for bringing  \cite{Craps:2016cgo} to our attention.}.
   
  \subsection{Deriving the equation}

  Here we derive the equation for all modes. We have seen that the ten dimensional modes have the scaling of operators with dimensions $\Delta$ given by \nref{OpDim}. From the two dimensional point of view (after reducing on the $S^8$) these are all scalar fields (though they can have spin under $SO(9)$).  When we lift them to $AdS_{2+\etanew }= AdS_{d+1}$ we expect to get scalar fields of mass 
  \be 
  m^2 R_{AdS}^2 = \Delta (\Delta -d)  
  \ee 
   In our case,  $d=14/5$, but this discussion is valid for general $d$ and $\Delta$. We then need to solve the wave equation for such a scalar field in the black brane geometry at finite temperature \nref{bbmet}. Writing the scalar field as $\Phi = e^{ - i \omega t } \chi(z) $,  we get 
    \be \label{zwe}
z^{d-1}\partial_{z}\left(\frac{h}{z^{d-1}}\partial_{z}\chi\right)+\left(\frac{\omega^{2}}{h} - \frac{\Delta(\Delta-d)}{z^{2}}\right)\chi = 0
    \ee 
    The boundary conditions at infinity are that the field behaves as $\chi \sim z^{\Delta}$ for small $z$. And at the horizon we impose the boundary condition
    \be \la{HorCon}
    \chi \sim (z_{0}-z)^{-\frac{ i z_{0} \omega}{d}} \sim \rho^{- \frac{i \omega  }{2 \pi T}}
    \ee 
    where $\rho$ is the proper distance from the horizon. 
    %This corresponds to a plane wave falling in to the future horizon. 
    There is another solution near the horizon behaving as in \nref{HorCon} but with $i \to -i$ in the exponent. \nref{HorCon} corresponds to the solution which is regular at the future horizon. Including time dependence, \nref{HorCon} becomes $ \Phi \sim (X^{+})^{-\frac{i \omega}{2 \pi T}}$ where $X^{+} =\rho  ~ \text{exp}(2 \pi T t)$ is one of the Kruskal coordinates in the near horizon region.
    %In that case we get something like $ (X^+)^{ -i\omega }$ where $X^+$ is one of the Kruskal coordinates in the near horizon region. 
    This expression is non-singular for finite $X^+$ and $X^-\to 0$ which is the future horizon. With these boundary conditions, there are solutions only for a discrete set of complex values of $\omega$, which are the quasinormal mode frequencies. 
   % The plus sign solution corresponds to a plane wave outgoing from the horizon and is irregular as $\rho \to 0$. Therefore this solution is discarded and we keep only the other, ingoing solution.  

Since our argument involved going to a fractional number of dimensions, as a sanity check,    we  also derived  the equation for the $SO(9)$ scalar mode by conventional methods. We expect a single $SO(9)$ singlet mode with dimension $\Delta = 28/5$, obeying \cite{Sekino:1999av}. 
We derived this by starting from the ten dimensional Schwarzschild black hole and lifting it to an eleven dimensional black string. In principle, we need to boost along the eleventh dimension $x_{11}$ in order to take the near horizon limit. Alternatively, we can consider only modes which in the final IIA picture will have no D0 brane charge. Those are the modes $\phi = e^{ - i \omega t + i k x_{11} }$ whose frequency and momentum along $x_{11}$ are related by $\omega = k$. We then expand the full metric in terms of $SO(9)$ scalar fluctuations, fix the gauge, etc, in the standard way. With this method we indeed get the wave equation in \cite{Sekino:1999av} at zero temperature. For non-zero temperature we get the same wave equation obtained above \nref{zwe}.  
%We give more details of this procedure in appendix XXX SHOULD WE DO THAT? \AB{I can add this if we decide it is worth including}

In the following, we set $z_0=1$ and in the end restore the temperature dependence from the formula 
    \be 
    T = { d \over 4 \pi z_0 } 
  \ee 
  
    We now tackle the mathematical problem of determining these modes.

 \subsection{Numerical results for the quasinormal mode frequencies}

QNM frequencies were computed using two independent numerical methods. In one method, we used a Mathematica package which implements a pseudospectral technique \cite{Jansen:2017oag, Jansen}. The package  replaces the radial variable of the wave equation by a grid $z_{i}$ of $n$ points and writes the solution $\chi(z)$ as a linear combination of polynomials, the so-called cardinal functions, each of which has support on a single gridpoint. This turns the wave equation into an $n \times n$ generalized eigenvalue problem, which is solved using standard linear algebra.

Boundary conditions are imposed implicitly by virtue of the cardinal functions used to approximate solutions. These functions are smooth and finite, and therefore cannot approximate solutions which are becoming arbitrarily large or oscillating arbitrarily quickly. Meanwhile, at the future horizon, the desired solution \nref{HorCon} is regular while the other solution is not. At the boundary, field redefinitions can be performed so that the solution which goes like $z^{\Delta}$ is normalizable while the solution we wish to set to zero (which goes like $z^{d-\Delta}$) is not. With this setup, the desired boundary conditions are automatically imposed.

%As mentioned previously, the outgoing mode oscillates more and more rapidly near the future horizon while the ingoing mode is perfectly smooth. 
%
%Furthermore, field redefinitions can be performed so that the solution which goes like $z^{\Delta}$ at infinity is normalizable while the other solution (which goes like $z^{d-\Delta}$) is not. Meanwhile, the cardinal functions are smooth and finite, and therefore cannot approximate functions which are becoming arbitrarily large or oscillating arbitrarily quickly. So the desired boundary conditions are automatically imposed.

Notably, an $n \times n$ eigenvalue equation generally yields $n$ eigenvalues. To see which eigenvalues correspond to QNM rather than numerical artifacts, we perform the same computation at various grid spacings and check for convergence.

In the second method, we construct Frobenius series solutions to the wave equation around the singular points at $z = 0$ and $z = 1$,  restricting to solutions which obey the desired boundary conditions
%solutions obeying the boundary conditions at those points 
% We expand the solutions with the right boundary conditions at $z=1$ and $z=0$ 
\bea \la{FrSe1}
\chi_{\text{hor}}(z) &\sim & (1-z)^{-i \omega/d} \left( 1 + \sum_{n\geq 1} a_n (z-1)^n \right) ~,~~~~~
\\  
\chi_{\text{bdy}}(z) &\sim&  z^{\Delta} \left( 1 + \sum_{n\geq 1} b_n z^{n/5} \right) \la{FrSe}
\eea
% \JM{I changed the power as you said, sorry about that.}
The coefficients $a_n$ and $b_n$ can be determined recursively using  equation \nref{zwe} and they depend on $\omega$. 
For general $\omega$ these define two independent solutions. We want to impose that they are proportional to each other. This can be ensured by setting their Wronskian to zero at some intermediate value of $z$. 
% Boundary conditions are specified by the exponent of the leading term: $\chi_{\text{hor}}(z) \sim (1-z)^{-i \omega/d}$ near $z = 1$ and $\chi_{\text{bdy}}(z) \sim z^{\Delta}$ near $z = 0$. 
The Wronskian of two solutions $\chi_{1}, \chi_{2}$ of (\ref{zwe}) is 
\begin{align}
W(\chi_{1}, \chi_{2}) = \frac{1-z^{d}}{z^{d-1}}\left(\chi_{1}\partial_{z}\chi_{2}-\chi_{2}\partial_{z}\chi_{1} \right)
\end{align}
$W$ is the conserved inner product associated to (\ref{zwe}). In other words, it is independent of the value of $z$ at which we evaluate it. Demanding that 
\be \la{QNMWro}
W(\chi_{\text{hor}}, \chi_{\text{bdy}}) = 0
\ee 
 at some intermediate $z_{0}$ within the radius of convergence for the Frobenius series in \nref{FrSe1} \nref{FrSe} yields an equation for $\omega$. We can get a numerical approximation for $\omega$ by truncating the sums in \nref{FrSe1} \nref{FrSe}.

\begin{figure}[!htbp]
   \begin{center}
   \includegraphics[height=10cm]{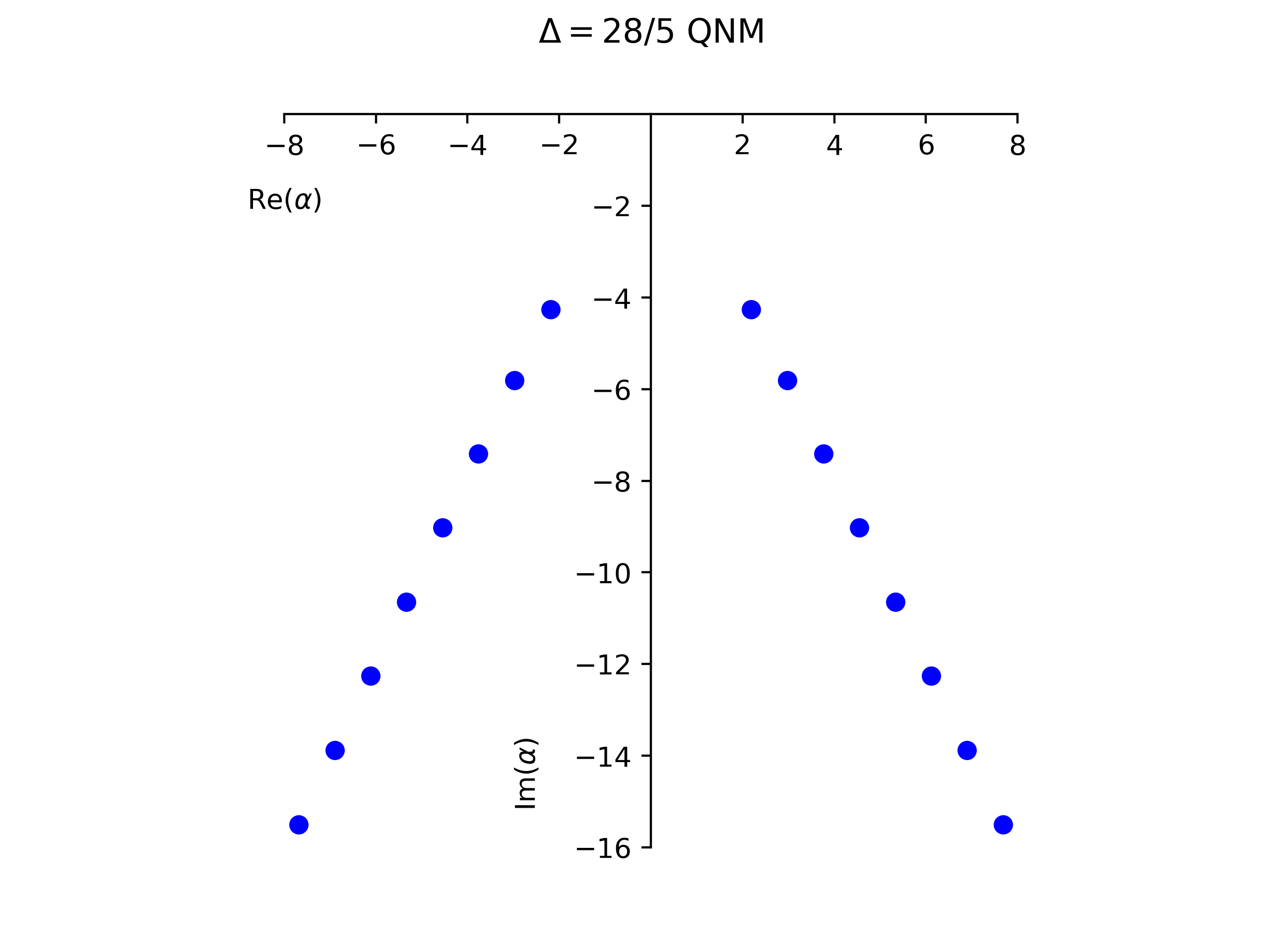}
    \end{center}
    \caption{ QNM frequencies for the $SO(9)$ scalar mode ($\Delta = 28/5$). }
    \label{k4l0plot}
\end{figure}

The first eight QNM frequencies for a sampling of fields are included in Tables 1-3 below. 
%We find a spectrum with the general $AdS$ form in all cases \JM{What does this sentence mean?}.
 An example of the typical spectrum for the lowest-lying modes is shown in Figure \ref{k4l0plot} for the $SO(9)$ scalar.

%\begin{center}\label{k4l0plot}
%\includegraphics[height=8.5cm]{k4l0QNM.png}
%\captionof{figure}{QNM frequencies for the $SO(9)$ scalar mode ($\Delta = 28/5$)}
%\end{center}
\begin{table}[!htbp]
    \begin{minipage}{.5\linewidth}
 \begin{center}
 \captionof{table}{$\Delta=4/5$ QNM}%\label{k0l2modes}
 \begin{tabular}{|p{0.3cm}|p{2.4cm}|p{2.4 cm}|}
 \hline
 $n$ & Re($\alpha_{n}$) & Im($\alpha_{n}$)\\
 \hline \hline 	
 1&$\pm 0.66509418$ & $- 2.21546231 $\\
 \hline
 2 & $\pm 1.31554955$ & $- 4.07444412$\\
 \hline
 3 &$\pm1.88708778 $ &$-5.93275042 $\\
 \hline
 4 & $\pm 2.43437862 $& $- 7.78935750 $\\
 \hline
 5 &$\pm 2.97124590 $ & $- 9.64490776 $\\
 \hline
 6 &$\pm 3.50268733  $ & $- 11.5001450 $\\
 \hline
 7 & $\pm 4.03082999 $&$- 13.3553942 $ \\
 \hline
 8 &$\pm 4.55675030$ & $- 15.2107465 $\\
 \hline
 \end{tabular}
 \end{center}
    \end{minipage}
    \begin{minipage}{.5\linewidth}
 \begin{center}
 \captionof{table}{$\Delta=14/5$ QNM}%\label{k0l7modes}
 \begin{tabular}{|p{0.3cm}|p{2.4cm}|p{2.4 cm}|}
 \hline
 $n$ & Re($\alpha_{n}$) & Im($\alpha_{n}$)\\
 \hline \hline 	
 1& $\pm 1.10811896$ & $-1.85999910$\\
 \hline
 2 & $\pm 1.89521921$ & $-3.48439968$\\
 \hline
 3 & $\pm 2.67872384$ &$-5.10830779$ \\
 \hline
 4 &$ \pm 3.46130872 $  &$-6.73200184$ \\
 \hline
 5 &$  \pm 4.24355190$ & $-8.35560521$\\
 \hline
 6 &$ \pm 5.02563617 $  & $-9.97916493$\\
 \hline
 7 & $ \pm 5.80763556 $ & $ -11.6027011 $ \\
 \hline
 8 & $ \pm 6.58958506 $ & $-13.2262233$ \\
 \hline
 \end{tabular}
 \end{center}
    \end{minipage} 
    \vspace{3mm}
    
   \begin{minipage}{ 1 \linewidth}
   	 \begin{center}
 \captionof{table}{$\Delta=28/5$ QNM}%\label{k4l0modes}
 \begin{tabular}{|p{0.3cm}|p{2.4cm}|p{2.4 cm}|}
 \hline
 $n$ & Re($\alpha_{n}$) & Im($\alpha_{n}$)\\
 \hline \hline 	
 1& $\pm 2.18097422 $ & $-4.25532498$\\
 \hline
 2 & $\pm 2.97857688$ & $-5.81267756$\\
 \hline
 3 & $\pm 3.76609055$ &$-7.41070529$ \\
 \hline
 4 &$ \pm 4.55050126 $  &$-9.02240909$ \\
 \hline
 5 &$  \pm 5.33377960$ & $-10.6397105$\\
 \hline
 6 &$ \pm 6.11654890 $  & $-12.2596427 $\\
 \hline
 7 & $ \pm 6.89904704 $ & $ -13.8809437$ \\
 \hline
 8 & $ \pm 7.68138146 $ & $-15.5030142$ \\
 \hline
 \end{tabular}
 \end{center}
   \end{minipage}
\end{table}

 \subsubsection{WKB approximations}

  For large mass we expect that the QNM will be given by a geodesic in the black brane background that is sitting at some value of $z$
 \be \la{ActQNM}
i S  = - i m R_{AdS}   \int d \tau  { 1 \over z } \sqrt{ 1 -z^d/z_0^d } 
\ee
Extremizing with respect to $z$ we get 
\be \la{zVal}
z = (-1)^{1/d } \left( { 2 \over d-2} \right)^{1/d} z_0 
\ee 
In principle we have several roots. We pick the ones with least imaginary part, namely $(-1)^{1/d} = e^{  i \pi/d}$. 
We can also expand the action for small fluctuations around this solution.  
Inserting \nref{zVal} into the action \nref{ActQNM} then gives us the QNM frequencies \cite{Festuccia:2008zx}  
\bea \label{WKBll}
 2 \pi T \alpha_n^{WKB} &\sim&   { 1 \over z_0} e^{ - i \pi/d}   \sqrt{d\over d-2}   \left( { d-2 \over 2} \right)^{1/d  }  \left(m R_{AdS} + \left( n + \half \right) \sqrt{d}  \right)
 \cr 
 &\sim &   { 1 \over z_0} e^{ - i \pi/d} \sqrt{d\over d-2}   \left( { d-2 \over 2} \right)^{1/d  }  \left( \Delta -{ d \over 2 }  + \left( n + \half\right) \sqrt{d}  \right)
 \eea 
 where $n=0,1,2, \cdots $ is an integer.   
 In the second expression we have inserted a more accurate expression derived in \cite{Festuccia:2008zx}. 
 Note that \nref{WKBll} has an overall factor of $e^{ - i \pi/d}$ which explains why the modes tend to lie on a straight line, see Figure \ref{k4l0plot}. They do not lie perfectly along such a line because \nref{WKBll} is just an approximation.
 
 In principle, \nref{WKBll} is valid only for large $\Delta$, but the formula works pretty well for the low lying values obtained above when $\Delta  \gtrsim d$. Writing the fractional error between a mode $\alpha$ and the WKB-predicted value  in (\ref{WKBll}) as
\begin{align}
 \epsilon_{n} = \bigg|{\alpha_{n} - \alpha_{n}^{WKB} \over \alpha_{n}}\bigg|
 \end{align}
we find that $\epsilon_{0} < 0.09$ when $\Delta \geq d$. The error decreases with $n$, or  as we move away from the fundamental mode toward higher frequencies. For the $\Delta = 28/5$ $SO(9)$ scalar, $\epsilon_{0} \sim 0.03$. Values of $\Delta$ below $d$ yield greater disagreement. For the lightest field in the spectrum, $\Delta = 4/5$, $\epsilon_{0} \sim 0.9$.

%
%For $\Delta = 28/5$, $\epsilon_{0} = 0.0290145$
%For $\Delta = 16/5, \epsilon_{0} =0.0721893$
%For $\Delta = 14/5, \epsilon_{0} =0.0883883$.
%For $\Delta = 4/5, 0.902785$.
%For $\Delta = 6/5, 0.757664$.
%For $\Delta = 8/5, 0.199657$.

As $n \to \infty$, we expect \cite{Festuccia:2008zx}  the QNM spacing to asymptote to 
 \be\label{WKBasy}
 \delta_{n}^{WKB} \equiv \alpha_{n} - \alpha_{n-1} =  \pm \frac{d}{2 \pi T} \sin \frac{\pi}{d} e^{\mp i \frac{\pi}{d}}
 \ee
We find that (\ref{WKBasy}) converges more quickly to the numerical results for larger $\Delta$. However, we expect (\ref{WKBasy}) also to hold for small $\Delta$, as long as $n$ is taken sufficiently large. For example, the percent error $|\delta_{n} - \delta_{n}^{WKB}| / |\delta_{n}|$ of (\ref{WKBasy}) decreases to 0.0008 by $n = 8$ for the $\Delta = 28/5$ $SO(9)$ scalar, while for the $\Delta = 4/5$ field, (\ref{WKBasy}) is still off by $18 \%$ at $n=8$.
 % \JM{Comment on the angle and the phase $e^{ - i \pi/d}$. } 
 % \AB{What do you have in mind for this? I noticed that for heavier fields, the WKB formula underestimates the angle between the two branches of modes, while for lighter fields it overestimates the angle. But I'm not sure this is particularly interesting. The phase also stops being correct for small $d$, as we discussed.}

\subsection{Sensitivity to the UV boundary conditions } 

The computation of quasinormal modes discussed in this section was done using gravity. Since the gravity description breaks down near the boundary, one could wonder about the sensitivity of the quasinormal modes to the precise boundary conditions\footnote{We thank G. Horowitz for this question.}. We can explore this as follows. Instead of putting the boundary condition at $z=0$, we   put it at $z=\epsilon$, where $\epsilon \propto T \lambda^{-1/3}$ in units where the horizon is at $z=1$. Then
the solution that obeys the modified boundary condition is 
\be 
\tilde \chi_{\rm bdy} = \chi_{\rm bdy} + \alpha \chi'_{\rm bdy} ~,~~~~~~~\alpha \propto \epsilon^{2\Delta -d} 
\ee 
where $\chi_{\rm bdy} $ is given in \nref{FrSe}, and $\chi'_{\rm bdy}$ is the other solution, with $\Delta \to d- \Delta $ in  \nref{FrSe}.  

The equation determining the quasinormal modes, analogous to \nref{QNMWro},  can now be expanded as 
\be \label{domega1}
0=W(\chi_{\rm hor} , \tilde \chi_{\rm bdy} ) = W(\chi_{\rm hor} ,  \chi_{\rm bdy} ) + \alpha W(\chi_{\rm hor} ,   \chi'_{\rm bdy} ) \propto \delta \omega + c \alpha ~~~\longrightarrow ~~ \delta \omega \propto \epsilon^{2\Delta -d}
\ee 
where $c$ is a constant  proportional to  $W(\chi_{\rm hor} ,   \chi'_{\rm bdy} )$ which is order one in $\epsilon$. We have also used that $W(\chi_{\rm hor} ,  \chi_{\rm bdy} )$ is proportional to $\delta \omega = \omega - \omega_n^0$, where $\omega_n^0$ is the quasinormal mode frequency when we select the original boundary conditions, see \nref{QNMWro}. 

\nref{domega1} describes the dependence of $\delta \omega$ on $\epsilon$. To determine the $\omega$ dependence, we consider a simpler problem, that of a massive scalar in empty $AdS_{d+1}$ with a hard wall at $z = 1$. Although the bulk boundary condition is different from the one we are interested in, we note that moving the UV boundary condition to $z = \epsilon$ only affects the form of the wavefunction at small values of $z$. Therefore we expect the $AdS$ hard wall problem to correctly capture the scaling of $\delta \omega$ even for the case where a black hole is present.

In empty $AdS$ the wave equation is solvable and the solutions are Bessel functions, $z^{d/2}J_{\pm (\Delta - d/2)}(\omega z)$. We first solve for the normal mode frequencies when Dirichlet boundary conditions are imposed at $z = 1$ and $z = 0$. We then shift the UV boundary condition to $z = \epsilon$ and solve for the resulting change $\delta \omega$ in the frequencies, using the asymptotic form of the Bessel functions $J_{\pm (\Delta - d/2)}(\omega \epsilon)$ near $\epsilon \to 0$.

This calculation suggests that $\delta \omega_n$ of the black hole QNM should scale like
\begin{align}\label{domega2}
\delta \omega_n/T \propto C\lp {\omega^{0}_n\over T}, \Delta, d\rp\frac{\Gamma(1-\Delta + \frac{d}{2})}{\Gamma(1+\Delta - \frac{d}{2})} \lp\frac{\omega_n^{0}}{2\lambda^{1/3}}\rp^{2 \Delta - d}  
\end{align}
where $C\lp{\omega^{0}_n \over T}, \Delta, d\rp$ is an order one factor that depends on the  details of the bulk solution and boundary condition. In restoring temperature dependence, $\epsilon \omega^{0}_n$ has been replaced by $\omega^0_n/\lambda^{1/3}$.  The scaling \nref{domega2} is supported by numerical analysis.

%up to an \omega, \delta, and d dependent factor that on the details of the bulk wavefunction and boundary condition
%the above formula does not 
%Setting $\chi|_{z = \epsilon} = 0$ and expanding the Bessel functions for small $z$ gives us the change in QNM frequencies
%\begin{align}\label{domega2}
%	\delta \omega \propto  \frac{\Gamma(1-\Delta + \frac{d}{2})}{\Gamma(1+\Delta - \frac{d}{2})}\lp \frac{\omega \epsilon}{2}\rp^{2 \Delta -d} \propto  \frac{\Gamma(1-\Delta + \frac{d}{2})}{\Gamma(1+\Delta - \frac{d}{2})}\lp \frac{\omega}{2 \lambda^{1/3}}\rp^{2 \Delta -d}
%\end{align}
%The scaling \nref{domega2} agrees with numerical analysis.

When $2\Delta -d > 0$  the quasinormal modes are not very sensitive to the precise boundary conditions as $\epsilon \to 0$. In fact, this change in boundary conditions can be interpreted as the insertion of a double trace operator $O_\Delta^2$ whose dimension is approximately $ 2\Delta$ \cite{Witten:2001ua}. When this is integrated over $d$ dimensions, this leads to effects that scale as 
$\epsilon^{2\Delta -d}$. In other words, when the double trace operator is irrelevant, the effects are small. 

On the other hand, if the double trace operator is relevant, $2\Delta -d <0$,  then the effects of a modified boundary condition is large. This is the case, for example, for the $\Delta = 4/5$ mode with QNM frequencies listed in Table 1. For such modes, we need a more sophisticated argument for selecting the boundary condition. For example, we could use the fact that these double trace operators typically break supersymmetry, so that the right boundary condition for the supersymmetry preserving model is such that supersymmetry is also preserved in the gravity approximation.

  \section{Other Dp brane geometries }

 In this section we generalize the discussion of the scaling similarity transformation for the near horizon geometries describing all $Dp$ branes \cite{Itzhaki:1998dd,Boonstra:1998mp}, see also \cite{Dong:2012se}. 
 
The general extremal geometry  is 
\bea 
ds^2 &=& f^{1/2} r^2 \left[  { dx^2_{p+1} \over r^2 f} + { dr^2 \over r^2 } + d\Omega_{8-p}^2 \right]
 ~,~~~~~~~f = {d_p g_{YM}^2 N \over r^{ 7-p} } 
 \cr 
 e^{ \phi } & \propto & g_{YM}^2 f^{ 3 - p \over  4 } ~,~~~~~~~~~A_{0\cdots p} \propto f^{-1} 
 \eea 
 
 This metric, and the corresponding action,  have a scaling similarity 
 \be 
 t \to \gamma t ~,~~~~~~~~ r\to \gamma^{ - { 2 \over 5 -p} } r ~,~~~~~~~ S \to \gamma^{-\etanew   }     S ~,~~~~~\etanew   = { (3-p)^2 \over 5-p } \la{EtaP}
 \ee 
% where we also indicated how the action transforms. 
 
 We can also formally add $\etanew  $ extra dimensions and uplift the solution to a solution which is 
 $AdS_{2+p + \etanew   } \times S^{8-p} $ \cite{Kanitscheider:2009as}, where the quotients of the radii are 
 \be \la{UpLI}
 {R_{AdS_{1+d}} \over R_{S^{8-p}} } = { 2 \over 5-p} = { d - 2 \over 2} ~,~~~~~~d = 1 +p + \etanew  
 \ee

 Note that once we {\it assume} that the gauge theory has a scaling similarity, then it is possible to compute the scaling exponent $\etanew  $ by a procedure similar to the one discussed in section 
 \ref{MMDer}. Namely, we consider configurations describing separated branes that are slow-moving in the separation.
 %, the slow velocity expansion. 
 We find that the vacuum expectation value of the matrix model field $X$ scales like $r \propto z^{ - { d-2\over 2} }$ in the above solution.   Using that the (velocity)$^4$ term has a protected form we can  fix the scaling exponent $\etanew  $ to \nref{EtaP}. 
 This constitutes an explanation for the peculiar powers of temperature that appear in the entropies of these solutions, 
 \be \la{NEent}
 S \propto N^2 V_p T^{ p +\etanew  } \lambda^{ -\etanew  /(3-p )}
 \ee 
 where we reintroduced  the 't Hooft coupling by dimensional analysis. 
 Note that probe actions also have this scaling similarity, which explains the agreement with the thermodynamics in the moduli space approximation   \cite{Wiseman:2013cda,Morita:2013wfa}. The logic here is different and it consists of deriving the temperature dependence in  \nref{NEent} from the gauge theory, specifically by making the assumption that we have a scaling similarity and then deriving the scaling exponent $\etanew  $ from first principles by using supersymmetry constraints on the low energy dynamics of multicenter solutions. 
 
 This argument also fixes the scaling dimensions of the simplest operators, $Tr[X^\ell]$,  to
 \be 
 \Delta = { (7-p )\over (5 -p)}  (b+2) + { 2 \over 5 -p} \ell  \la{SpeP}
 \ee 
 with $b=-2$. Other operators have similar dimensions but with  $b =-1,0,1,2$.    
These are the dimensions for massless (bosonic) supergravity fields. For fermions we expect half integer values of $b$. For massive string states we expect a discussion similar to the one in section \ref{MassStr} but with $3/5 \to (3-p)/(5-p)$ in \nref{massz} and $3/10 \to  \half (3-p)/(5-p)$ in \nref{Acz} and \nref{CorrMS}.

 The finite temperature solution uplifts to the usual $AdS_{1+d}$ black brane.  The equation for quasinormal modes can then be easily written as a perturbation of those black brane solutions, using the spectrum of fields in \nref{SpeP}, and has the form \nref{zwe} with $d$ and $\Delta$ given by  \nref{UpLI} \nref{SpeP}. 
 
 There is an interesting suggestion in \cite{Sekino:2019krf,Kitamura:2021adx} for how to go from the strong coupling dimensions  of the operators $Tr[X^\ell]$ in \nref{SpeP} to the weak coupling dimensions corresponding to $\ell $ free fields in $p+1$ dimensions by taking a suitable large $\ell$ limit similar to the one in \cite{Berenstein:2002jq}. %\JM{Added reference to Sekino}

 Let us discuss some special cases now. 
 
 \begin{itemize}
 	\item $p=-1$: In this case $\etanew   = { 8 \over 3} $. One can wonder about the meaning of the exponent in this case. One observation is the following. Consider the fermion mass operators  as in   \nref{FerMass},  which for general $p$ have  a dimension such   that its coefficient $\zeta$ has mass dimension one, call it $\zeta = \mu$. This is a relevant deformation around the critical point.    Then if we perturb the action by   such operators,  we predict that its action should scale as $S \propto  \mu^{8/3}$. This is a non-trivial prediction of the scaling similarity\footnote{For a particular choice of operator this integral was computed in \cite{Moore:1998et}. In that case the coefficient of the classical action term is zero, because the operator preserves some supersymmetries. We expect that the coefficient in the classical action should be non-zero for operators that break enough supersymmetries. In making this prediction we also assumed that the solution only has the $F_9$ RR flux as a quantized flux, otherwise the scaling similarity would interfere with the flux quantization condition. }.   Note that we are saying that the matrix integral develops a non-trivial critical point which can be seen by looking at the values of the integral upon adding some deformations\footnote{Scale invariance and matrix integrals were recently explored in \cite{Aguilar-Gutierrez:2022kvk} in a related context.}. 
 	\item $p=1$: Here $\etanew  =1$. we can think of the solution as S-dual to that of the fundamental string in type IIB, and this solution is similar in type IIA. In type IIA we can uplift the solution to the usual M2 brane solution or $AdS_4 \times S^7$ which are the dimensions that also appear here for that case. 
 	\item $p=2$: We get $\etanew   = 1/3$, which is a non-trivial similarity exponent. Notice that in the far IR the boundary quantum field theory develops an actual conformal fixed point which is  dual to $AdS_4 \times S^7$. This  is {\it different} that the one we get in the intermediate energy regime described here. 
 	\item $p=3$: We get $\etanew  =0$ and the usual CFT. In this case the two scaling similarities we discussed in \nref{nuSym} \nref{TenRes} combine to imply that the gravity answers are independent of the string coupling and have an action proportional to $N^2$, which is of course a familiar result. 
 	\item $p=4$: We get $\etanew   =1$ and the uplift is $AdS_{7} \times S^4$. This is just saying that the metric is simply a dimensional reduction of the familiar solution for M5 branes. 
 	\item $p=5$: Here $\etanew  =\infty$ and the above formulas become singular. The proper interpretation is that the coordinates are not really rescaled but the radial direction,  $r$, is rescaled,   which is what happens when we go to the NS 5brane geometry. The similarity just corresponds to a shift of the dilaton as we shift  the radial dimension. 
     \item $p=6$: Here $\etanew   =-9$. The D6 brane solution can be uplifted to eleven dimensions as an $A_{N-1}$ singularity and the rescaling is just the usual rescaling of all coordinates in eleven dimensions. This gives the usual scaling of the action as (length)$^9$, which is the standard similarity in eleven dimensional flat space. 
 	 \end{itemize}

  {\bf Note added}\footnote{We thank N. Bobev, P. Bomans and F. Gautason for bringing  \cite{Bobev:2019bvq} to our attention,  and for some comments that gave rise to this added note.} 
  
  In the interesting paper \cite{Bobev:2019bvq}, the sphere partition function was computed for the gauge theories living on the worldvolume of $Dp$ branes.   Using supersymmetric localization they computed the logarithm of the partition function on $S^{p+1}$. For $p \geq 1$ they found that it scales as $(\bar \lambda)^{p-3 \over 5-p}$ in terms of the dimensionless coupling  $\bar \lambda = { \lambda R^{ 3-p}} $, where $R$ is the radius of the sphere (see eqn. (2.19) in \cite{Bobev:2019bvq}). This is precisely the scaling that we expect on the basis of the scaling similarity, namely the total power of size of the sphere is then $R^{ - \etanew }$. In fact, this observation can be turned around and viewed as an argument for a computation of $\etanew$ from the matrix model, logically similar to what we discussed in section \ref{MMDer}. Namely, we assume that the theory has a scaling similarity and we used the localization results to compute the exponent.  Of course, the power law dependence on $\bar \lambda$ observed in \cite{Bobev:2019bvq} is also evidence that the matrix model has this similarity.   
    
   In \cite{Bobev:2019bvq},  the Wilson loop was also computed. This is also constrained by the scaling similarity, but with a different action scaling exponent than the one we have for the bulk gravity action. The relevant exponent is the one that appears in the action of the string, which is related to the scaling of the string frame metric, which goes as $ds^2_{str} \propto z^{ 3-p \over 5-p } \left[ AdS_{2 +p } \times S_{8-p} \right]$, where the metric of $AdS_2$ is $ { dt^2 + dz^2 \over z^2 } $. This implies that the scaling exponent for the fundamental string action is $\hat \theta = { p -3 \over 5-p}$. Therefore the Wilson loop action is expected to scale as $(\bar \lambda)^{1/(5-p)}$, which is indeed what \cite{Bobev:2019bvq} found in their equations (2.15) and (2.21)-(2.22).

%    \textcolor{purple}{While looking at \cite{Bobev:2019bvq} I was momentarily confused by their equation (1.1), where they write that the dimensionless coupling is $\bar \lambda = \lambda R^{p-3}$, which is different than what is written in the above note. However I think their equation must be wrong by dimensional analysis. - AB}. \JM{ Yes, their equation must be wrong!}
   
 \section{Discussion} 
 
 In this paper we discussed in some detail the scaling similarity of the $D0$ brane solution. We emphasized that it should be viewed as a non-trivial scaling transformation which is not obvious from the matrix model point of view, and it would be interesting to explain its emergence purely from the matrix model.  On the other hand, the assumption that the matrix model develops a similarity, plus the constraints of supersymmetry, determine the important action scaling exponent $\etanew  $. In turn,  this determines how the finite temperature entropy depends on the temperature.  
% We emphasized that we assume a similarity with a general action scaling exponent $\etanew  $ and we fix $\etanew  $ from a purely matrix mode. 
 Similar similarities exist for other $Dp$ brane solutions (of course for $p=3$ this is actually an ordinary scaling symmetry). 
 %We also determined the action scaling exponent $\etanew  $ from the boundary gauge theory point of view by using supersymmetry constraints on brane probe actions. 
%\AB{I commented out two sentences in the above paragraph that I thought seemed redundant.}

Here we noted that a particular many-body quantum system, the matrix model, develops a scaling similarity in the large $N$ limit. It would be interesting to find other examples of this phenomenon. In particular one could wonder whether there is an SYK-like system that exhibits a scaling similarity with a non-zero action scaling exponent $\etanew  $. In statistical mechanics,  an example is the random field Ising model \cite{Fisher:1986zz}. In \cite{Lin:2013jra} some supersymmetric quantum mechanics models were analyzed by solving the one loop truncated\footnote{Though this truncation is an uncontrolled approximation, it is interesting that they found a scaling behavior. }    Dyson-Schwinger equations and they found a scaling similarity.  

 We have also used the eleven dimensional uplift of the D0 brane geometry to get the scaling dimensions of all gravity fields. A similar analysis can be done for $Dp$ branes by starting with a plane wave geometry smeared along $p$ dimensions and then performing U-dualities. This also leads to \nref{SpeP}.

  We then explored the quasinormal modes. At first sight, this is a difficult task because we need to expand all fields around the finite temperature solution. This task is simplified by a mathematical trick, uplifting the solution to $10+\etanew  $ dimensions, so that we have an $AdS_{1+d} \times S^{8-p}$ solution with $d= 1+ p + \etanew  $. From the $AdS_{1+d}$ solution point of view we have fields whose dimensions are given by \nref{SpeP} and we can simply write down the wave equation. 
  
  We analyzed this equation numerically for the D0 brane case, $p=0$. The quasinormal modes are a unique ``fingerprint'' of the black hole and it would be interesting to reproduce them from the matrix model by either a classical numerical simulation or by a quantum simulation.    
   An interesting feature of these modes is that the single trace operators come in special low spin representation of $SO(9)$, essentially spin less than two. It is not obvious from the matrix model that this is the case, but it is obvious from the bulk, since spins are upper bounded by two in gravity. The fact that we only get these light modes is analogous to the spin gap condition that was discussed in \cite{Heemskerk:2009pn} as a necessary (and probably sufficient \cite{Caron-Huot:2021enk}) condition for a gravity description.

 %\section{References on hyperscaling }
 
 %\cite{Huijse:2011ef}  defines $\theta = -\etanew  $ and refers to \cite{Fisher:1986zz} who define this exponent as hyperscaling violating exponent. $t \to \zeta t$, $ds \to \zeta^{\theta/d} ds$ with 
% $d$ the number of spatial dimensions, the dimension of the bulk is $D= d+2$ (one more radial and time dimensions). Massive field correlators behaves as $e^{ - x^{\theta/d} }$. 
 
% See also \cite{Dong:2012se}. They also cite \cite{Perlmutter:2010qu}
    
%   \section{Understand}
%   Understannd \cite{Jevicki:1998yr} on the probe action and conformal symmetry.. 
    
\subsection*{Acknowledgments}

We would like to thank M. Green, M. Ivanov, V. Ivo, M. Hanada,  M. Mezei,  M. Rangamani and J. Santos  for discussions.  We also thank  N. Bobev, P. Bomans, F. Gautason, G. Horowitz, R. Monten, K\'evin Nguyen and T. Wiseman for comments that we incorporated in the revised version. 

J.M. is supported in part by U.S. Department of Energy grant DE-SC0009988.

\appendix

\section{Details on the dimensional reduction} 
\la{Reduction}

The following formulas are useful for the dimensional reduction and uplift. See also \cite{Kanitscheider:2009as}.  

Starting from the Einstein action in $10+\etanew  $ dimensions $I_{10+\etanew  } = \int d^{10+ \etanew   }x \sqrt{\hat g } \hat R - { 1\over 2 }{ 1 \over  8!}F_8^2$ we 
write the metric as $d\hat s^2 = d\tilde s^2_{10} + e^{2 \sigma } d \vec x_\etanew  ^2 $. Then the curvature is 
\be 
\hat R = \tilde R - 2 \etanew   \tilde \nabla^2 \sigma - \etanew   (\etanew   +1) (\tilde \nabla \sigma )^2 
\ee 
The action becomes $I_{10+\etanew   } \propto  \int d^{10} x \sqrt{\tilde g } e^{\etanew   \sigma} \left[ \tilde R + \etanew   (\etanew  -1) (\tilde \nabla \sigma )^2 - { 1\over 2 }{ 1 \over  8!}F_8^2  \right] $ after an integration by parts. 
%\AB{Here there is an overall factor of $\text{Vol}(T^{\etanew})$ from the $\int d^{\etanew}x$ that has been dropped, right? Maybe it would be clearer to write $I_{10 + \etanew} \propto...$} 
To connect with the string metric we write $\tilde g_{\mu \nu}  = e^{ a \sigma } g_{\mu \nu}$. We then choose $a$ so that there is no $\sigma$ dependence in front of $F_8^2$. This sets $a = \etanew  /3$. 
We also use the formula for the form of the curvature under a rescaling of the metric 
\be 
\tilde R = e^{ - a \sigma } \left[ R - 9 a \nabla^2 \sigma - 18 a^2 (\nabla \sigma)^2 \right]
\ee 
Inserting this into the action we get that 
\be 
I_{10+ \etanew   } = \int \sqrt{ g} e^{ 7 a \sigma } \left\{ R + { (54 a^2 - 3 a) \over 49 a^2 } [ \nabla (7 a \sigma) ]^2 \right\} - { 1\over 2 }{ 1 \over  8!} F_8^2 
\ee  
where we used that $a = \etanew  /3$.
Setting this to be equal to the usual ten dimensional action, $ I_{10} = \int d^{10}x \sqrt{g} e^{ - 2 \phi } [ R + 4 (\nabla \phi)^2 ] - { 1\over 2 }{ 1 \over  8!} F_8^2 $,  implies \nref{etaphi}.  
    
 We can follow the same procedure for $Dp$ branes. In that case one has an $F_{8-p}$, so we get that 
 $a = \etanew  /(3-p)$, $\phi = - (7-p) \sigma/2$, $\etanew   = (3-p)^2/(5-p)$, and $d = 1 + p + \etanew   = 2 (7-p)/(5-p)$.

As a side comment,  note that a gravity solution of the form $AdS_{D} \times S_{D'}$ with an $F_{D'}$ field on the $S_{D'}$ (or its dual) will have radii in the following ratio
\be 
{ R_{S_{D'}} \over R_{AdS_{D} } } = { D' -1 \over D-1} 
\ee

   \section{Derivation of the one loop eight derivative correction}
   
   \la{RFour}
   
   In this appendix we provide a short  derivation of the one loop $R^4$ correction that was obtained originally in 
   \cite{Hyakutake:2013vwa,Hyakutake:2014maa}, and was matched to numerical results in \cite{Hanada:2013rga,Berkowitz:2016jlq,Pateloudis:2022ijr}.

  The ten dimensional effective action has higher derivative corrections which lead to a modification of the thermodynamic properties of black holes. These are particularly interesting because they can be compared with the numerical results in \cite{Berkowitz:2016jlq,Pateloudis:2022ijr}. 
  The first corrections are  eight derivative, or $\alpha'^3$,  corrections that arise from two sources. There is a tree level correction and a loop level correction of the schematic form 
  \be \la{CTA}
  - I = { 1 \over 16 \pi G_N} \int \sqrt{g} R  +  { 1\over \alpha' } \int  \sqrt{g} \left[ \# e^{ - 2 \phi } ( R^4 + \cdots ) + \#' ( R^4 + \cdots ) \right] 
  \ee 
  where the first term arises at string tree level and the second at one loop. The dots include terms involving the RR fields and gradients of the dilaton. The full form of the tree level term is not known. However, the one loop term is believed to be given simply by a Kaluza Klein reduction from a similar $R^4$ term in eleven dimensions. 
  %It leads to a correction to the thermodynamic properties of black holes and was originally computed in \cite{Hyakutake:2013vwa,Hyakutake:2014maa}.  
  
 The eleven dimensional correction to the Euclidean action is \cite{Green:1997as,Tseytlin:2000sf}
 \bea 
 - I &=& { 1 \over 16 \pi G_{N,11} } \int \sqrt{g} R + { \gamma \over l_p^3} \int \sqrt{g} J_M
 \\ 
 ~& ~& 16 \pi  G_{N,11} = (2 \pi)^8 l_p^9 ~,~~~~~~~~~\gamma = { 1 \over 2^{19} \times 3^2 \times \pi^6 } \la{GaDe}
 \eea 
 with 
 \be 
 J_M = t_8t_8 R^4 + { 1 \over 4 } E_8
 \ee 
 where the invariants $t_{8}t_8 R^4$ and $E_{8}$ can be expressed in terms of the Riemann tensor as in \cite{deRoo:1992zp} (with $E_8 = 8! Z$, with $Z$ in \cite{deRoo:1992zp}).  
 
Since the eleven dimensional solution \nref{11dmet} is simply a boosted version of a ten dimensional Schwarzschild black hole, we can simply evaluate the invariants on the ten dimensional Euclidean Schwarzschild black hole of the form 
\be 
 { ds^2 \over l_p^2}   =  f d\hat \tau^2 + { dy^2 \over f} + y^2 d\Omega_{8}^2 + dx_{11}^2 ~,~~~~~~~~~~~~f = 1- { y_0^7 \over y^7} 
 \ee 
 The non-trivial components of the Riemman tensor are 
 \be 
 R_{\hat 1 \hat 2 \hat 1 \hat 2  }  = 28 \mu ~,~~~~~~~
 R_{\hat 1 i \hat 1 j  } = R_{\hat 2 i \hat 2 j  } =- { 7 \over 2 } \delta_{ij} \mu ~,~~~~~~~~~ R_{\hat i \hat j \hat i \hat j } =  \mu ~~( {\rm{no ~sum}}, ~ i\not = j) ~,~~~~\mu \equiv { y_0^7 \over y^9} ~~~~
 \ee 
 where we gave the values in local Lorentz indices, with  $\hat 1 = \hat {\hat \tau}$,  $\hat 2 = \hat y $, and $i,j$ run over the eight values corresponding to the sphere directions. All other components not related to those above by the symmetries of the Riemann tensor are zero. 
 
 We can compute the various invariants for this black hole: 
\bea \la{SiDe}
 J_M &=& t_8t_8 R^4 + { 1 \over 4 } E_8 =\sigma \mu^4 ~,~~~~~~~~\sigma \equiv 2^{10} \times 3^5 \times 5 \times 7 \times 61
 \cr 
 t_8t_8 R^4 &=& 
  2^9 \times 3^5 \times 5 \times 7 \times 109 \, \mu^4
 \cr 
 {1 \over 4 } E_8 &=&
  2^9 \times 3^5 \times 5 \times 7 \times 13\, \mu^4
 \cr \la{siprime}
 J_0 &=& t_8t_8 R^4 - { 1 \over 4 } E_8  =\sigma' \mu^4~,~~~~~~~\sigma' \equiv 2^{14} \times 3^6 \times 5 \times 7   
 \la{FInJz}
 \eea
  where we found the formulas in \cite{deRoo:1992zp} useful. We have given the separate values of other invariants just for completeness. 
  
We can now evaluate the correction to the free energy by integrating $J_M$ over the whole space. Note that we do not need to find the correction to the black hole metric, since that would only change the answer at higher order. 
We obtain 
\bea
\delta \log Z &=& - \delta I = \gamma \omega_8 (2 \pi) \beta_{\hat \tau}  \sigma \int_{y_0}^\infty dy y^8 \left( { y_0^7 \over y^9} \right)^4 =  \gamma \sigma \omega_8 2 \pi \beta_{\hat \tau } y_0 { 1 \over 27} 
\eea
where $\omega_8 = 32\pi^4/105$ is the volume of the eight-sphere. 
 The factor of $2\pi$ is from the size of the $11^{\text{th}}$ direction, and $  \beta_{\hat \tau}$ is the size of the circle $\hat \tau = \hat \tau +   \beta_{\hat \tau}$, with  
\be 
\beta_{\hat \tau } = { 4 \pi y_0 \over 7 } \sqrt{ \alpha \over y_0^7}  ~,~~~~~~~\beta_{\hat \tau} = (2 \pi g)^{2/3} \beta_t 
\ee 
where $\beta_t$ is the physical temperature with respect to the time in the matrix model, and $ \beta_{\hat \tau}$ is the temperature, properly computed using \nref{11dmet}.  
  
Plugging in all those numbers we get 
\be \la{OneLoop}
\delta \log Z =
{ 61 \times 5^{1/5} \pi^{2/5} \over 2 \times 3^{4/5} \times (14)^{2/5} } 
( \beta \lambda^{1/3} )^{3/5} = 9.612 \left( { \lambda^{1/3} \over T } \right)^{3/5} ~,~~~~~~~~~\lambda \equiv g^2 N
\ee 
Then the correction to the energy   $E=-\partial_\beta \log Z $ is given by
\be 
\delta E = - { 3 \over 5} { 61 \times 5^{1/5} \pi^{2/5} \over 2 \times 3^{4/5} \times (14)^{2/5} } 
( \beta \lambda^{1/3} )^{3/5} { 1 \over \beta} = - 5.76709 \, \,  T^{2/5} ~,~~~~~{\rm for } ~~~\lambda =1 \ee 
as was obtained in  \cite{Hyakutake:2014maa} \cite{Hyakutake:2013vwa}. 

 Note that the one loop correction \nref{OneLoop} is larger than naively expected from a scaling point of view. Namely, from the gravity side, one might naively expect that a one loop correction is the logarithm of a determinant which has at most a logarithmic scaling with the scale or the temperature,  $\log ( \lambda^{1/3}/T) $.    On the other hand, \nref{OneLoop} is a correction that grows as a power when we reduce the temperature. The reason is that this correction comes from a one loop counterterm to \nref{CTA}, which scales as $({\rm size})^2/\alpha' \propto T^{-3/5}$, where the size refers to the radius of the $S^8$ in string units at the horizon, see   \nref{SolIIA}. This makes the correction larger than what we would expect if the gravity theory were one loop finite. Of course, the gravity theory is not finite, but the string theory is indeed finite, however the final answer is larger than naively expected and it depends explicitly on the string theory length $\alpha'$.  
  
  Just for completeness we also quote the leading order partition function 
  \be 
  \log Z = - \beta F = -I = N^2 T^{ 9/5} { 2^{21/5} 3^{2/5} 5^{7/5} \pi^{14/5} \over 7^{ 19/5} } =  N^2 T^{ 9/5} 4.115\cdots  ~,~~~~~{\rm for } ~~\lambda =1
  \ee 
   
  There is a numerical prediction for the tree level correction to the energy (of order $ \delta E = ( - 10.0 \pm 0.4)  N^2 T^{23/5}$) in \cite{Berkowitz:2016jlq,Pateloudis:2022ijr} which would be nice to verify. It would involve finding the full tree level $\alpha'^3$ correction involving the dilaton and the RR two form field strength. 

\eject

\bibliographystyle{apsrev4-1long}
\bibliography{GeneralBibliography.bib}
\end{document}